\RequirePackage{lineno}
\pdfoutput=1
\documentclass[preprint,prd,tightenlines, superscriptaddress]{revtex4}

\usepackage{graphicx} 
\usepackage{dcolumn}  
\usepackage{colordvi}
\usepackage{color}
\usepackage{epstopdf}
\usepackage{subfig}
\usepackage{caption}
\usepackage{amssymb}
\usepackage{amsmath}
\usepackage{url}
\graphicspath{{ps}}
\usepackage{hyperref}
\usepackage{tabularx}
\usepackage{comment}
\usepackage{booktabs}

\renewcommand{\arraystretch}{1.1}

\input belle2sym.tex

\def\Bsig        {\ensuremath{B_{\mathrm{sig}}}\xspace}

\def\X           {\ensuremath{X}\xspace}

\def \Y       {\ensuremath{\Upsilon(4S)}\xspace}

\begin{document}


\def\belletwo {{Belle II}\xspace}
\def\itbelletwo {{\it {Belle II}}\xspace}
\def\phaseiii {{Phase III}\xspace}
\def\itphaseiii {{\it {Phase III}}\xspace}

\newcommand\logten{\ensuremath{\log_{10}\;}}

\newcommand\bfrhoc{$\mathcal{B}$($B^0\to\rho^-\ell^+\nu_\ell$)}
\newcommand\bfrhoz{$\mathcal{B}$($B^+\to\rho^0\ell^+\nu_\ell$)}
\newcommand\bfpic{$\mathcal{B}$($B^0\to\pi^-\ell^+\nu_\ell$)}
\newcommand\bfpiz{$\mathcal{B}$($B^+\to\pi^0\ell^+\nu_\ell$)}


\def\lint {62.8 \invfb}
\def\procversion {{\it {proc11 + prompt}}\xspace}
\def\mcversion {{MC13a}\xspace}
\def\release {{\it {light-2002-ichep}}\xspace}
\def\feitraining {{\it {FEIv4\_2020\_MC13\_release\_04\_01\_01}}\xspace}
\def\sigprobcut {{$0.001$}\xspace}


\vspace*{-3\baselineskip}
\resizebox{!}{3cm}{\includegraphics{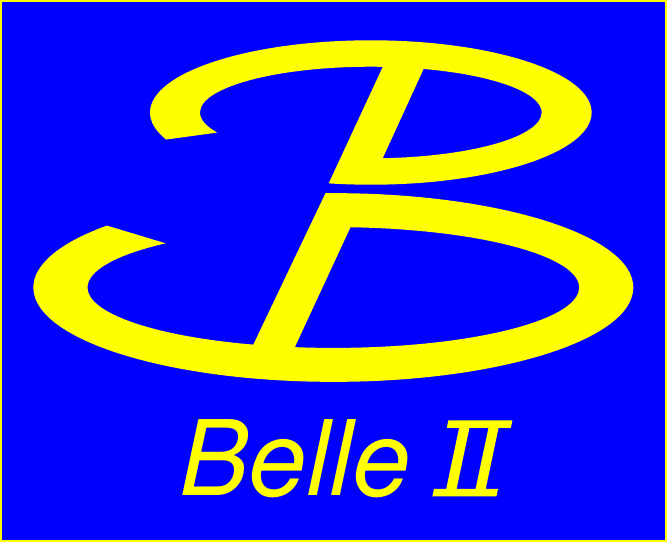}}

\vspace*{-5\baselineskip}
\begin{flushright}
\vspace{1em}
BELLE2-CONF-PH-2022-008\\
June 16, 2022

\end{flushright}

\vspace{2em}
\title {\LARGE Study of Exclusive $B \to \pi e^+ \nu_e$ Decays with Hadronic Full-event-interpretation Tagging in 189.3\invfb of Belle II Data\\
\vspace{1em}
}

\author{F. Abudin{\'e}n, I. Adachi, K. Adamczyk, L. Aggarwal, P. Ahlburg, H. Ahmed, J. K. Ahn, H. Aihara, N. Akopov, A. Aloisio
% ...
}

\collaboration{\textit{Belle II Collaboration}}





\begin{abstract}

We present a reconstruction of the semileptonic decays $B^0 \to \pi^- e^+ \nu_e$ and $B^+ \to \pi^0 e^+ \nu_e$ in a sample corresponding to 189.3\invfb of Belle II data, using events where the partner $B$-meson is reconstructed from a large variety of hadronic channels via a tagging algorithm known as the full-event-interpretation. We determine the partial branching fractions in three bins of the squared momentum transfer to the leptonic system using fits to the distribution of the square of the missing mass. The partial branching fractions are summed to determine $\mathcal{B}(B^0 \to \pi^- e^+ \nu_e)$ = (1.43 $\pm$ 0.27(stat) $\pm$ 0.07(syst)) $\times 10^{-4}$ and $\mathcal{B}(B^+ \to \pi^0 e^+ \nu_e)$ = (8.33 $\pm$ 1.67(stat) $\pm$ 0.55(syst)) $\times 10^{-5}$. We extract a first Belle II measurement of the magnitude of the Cabibbo-Kobayashi-Maskawa matrix element $|V_{\mathrm{ub}}|$, with $|V_{\mathrm{ub}}|$ = (3.88 $\pm$ 0.45) $\times 10^{-3}$.


\end{abstract}


\maketitle

{\renewcommand{\thefootnote}{\fnsymbol{footnote}}}
\setcounter{footnote}{0}


\section{Introduction}

Since the start of its first physics operation in 2019, the Belle II detector has collected over 350\invfb of data from electron-positron collisions. These early data have been invaluable for investigating the performance of the detector and the analysis software.

In this document, we present a reconstruction of the decays $B^0 \to \pi^- e^+ \nu_e$ and $B^+ \to \pi^0 e^+ \nu_e$, \footnote{Charge conjugate processes are implied for all quoted decays of $B$-mesons throughout this paper.} in a sample corresponding to 189.3\invfb of Belle II data. These decays are considered golden modes for precise determinations of the magnitude of the Cabibbo-Kobayashi-Maskawa (CKM) matrix element $|V_{\mathrm{ub}}|$. The reconstruction is performed via hadronic $B$-tagging provided by the full-event-interpretation (FEI) algorithm \cite{Keck:2018lcd}. While the integrated luminosity collected at present is too small to provide a competitive measurement compared with the current world average \cite{Zyla:2020zbs}, we demonstrate the first extraction of $|V_{\mathrm{ub}}|$ via a hadronically tagged approach at Belle II. The results presented build upon prior measurements made at an integrated luminosity of 62.8\invfb \cite{old:pilnu}.

\section{The Belle II Detector}
The Belle II detector is described in detail in Ref. \cite{Abe:2010sj}. The innermost layers are collectively known as the vertex detector (VXD), and are dedicated to the tracking of charged particles and the precise determination of particle decay vertices. The VXD is composed of two layers of silicon pixel sensors surrounded by four layers of silicon strip detectors. The central drift chamber (CDC) surrounds the VXD, encompassing the barrel region of the detector, and is responsible for the reconstruction of charged particles and the determination of their momenta and electric charge. The CDC additionally plays an important role in particle identification, with the use of specific ionisation information of charged particles.

Particle identification is also provided by two independent Cherenkov-imaging instruments, the time-of-propagation
counter and the aerogel ring-imaging Cherenkov detector, located in the barrel and forward endcap regions of the detector, respectively. The electromagnetic calorimeter (ECL) encases all of the previous layers and is primarily used for the determination of the energies of charged and neutral particles. A superconducting solenoid surrounds the inner components and provides the 1.5 T magnetic field required by the VXD and CDC. Finally, the $K^0_L$- and muon detector forms the outermost detector layer aimed at the detection of $K^0_L$ mesons and muons.

\section{Data sets}
\label{sec:data sets}

The amount of data studied for this analysis corresponds to an integrated luminosity of 189.3\invfb. To understand the properties of signal and background, fully simulated Monte Carlo (MC) samples of decays of pairs of charged or neutral $B$ mesons, as well as 
continuum \epem \to \qqbar ($q = u,\,d,\,s,\,c$) processes are used, corresponding to a total integrated luminosity of 400\invfb. These samples are generated alongside beam background effects including beam scattering and radiative processes. 

In addition to these MC samples, dedicated samples of $B \to \X_u \ell \nu_\ell$ decays, where $X_u$ is a hadronic system resulting from the $b \to u$ quark flavor transition  and $\ell = e$ or $\mu$, are used to model signal decays and related backgrounds. The $X_u$ system includes both resonant and nonresonant contributions using the hybrid modelling technique of Ref. \cite{Ramirez:1990db}, which is briefly described here.

Each $B^+ \to \X_u \ell \nu_\ell$ and $B^0 \to \X_u \ell \nu_\ell$ sample consists of a total of $50 \times 10^{6}$ resonant (R) events containing the relevant exclusive decays as well as $50 \times 10^{6}$ nonresonant (I) events corresponding to the inclusive component, simulated using the BLNP heavy-quark-effective-theory-based model \cite{Lange:2005ll}. These samples are then combined together and the eFFORT tool \cite{markus_prim_2020_3965699} is used to calculate an event-by-event weight $w_i$ in three-dimensional bins $i$ of the generated lepton energy in the $B$-frame, $E^B_\ell$, the squared four-momentum transfer to the leptonic system, $q^2$, and the mass of the hadronic system containing an up-quark, $M_X$, such that $H_i = R_i + w_i I_i.$ The number of total hybrid events per bin, $H_i$, is the sum of the number of resonant events $R_i$ and the number of inclusive events $I_i$ scaled by the appropriate weight $w_i$. 

The $B \to \X_u \ell \nu_\ell$ events from the charged and neutral $B$-meson decays in the 400\invfb MC sample are replaced with the equivalent amount of this hybrid re-weighted MC simulation.

\section{full event interpretation}
\label{sec:FEISkim}
\begin{figure}[h!]
\begin{center}
\includegraphics[scale=0.8]{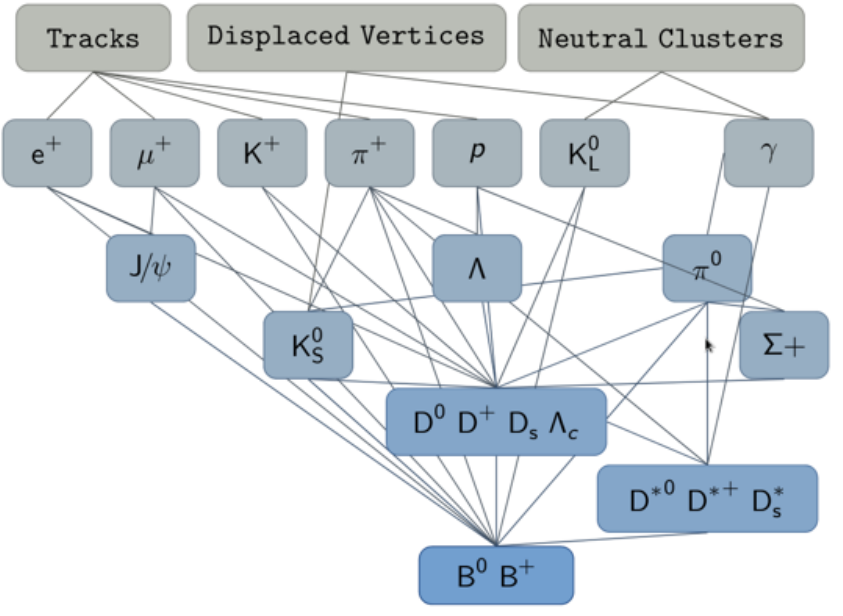}
  \caption{Hierarchical structure of the full-event-interpretation tagging algorithm. 
    }
  \label{fig:fei}
\end{center}
\end{figure}

The second $B$-meson in the $B\overline{B}$ pair is reconstructed using the full-event-interpretation (FEI) method \cite{Keck:2018lcd} to identify if the collision produced a $B\bar{B}$ pair (tag the event).
The FEI method is based on a machine learning algorithm developed for $B$-tagged analysis at Belle II. It identifies $B\bar{B}$ decays in both semileptonic and hadronic final states, reconstructing $B$ mesons across more than 4000 individual decay chains. The algorithm utilises a FastBDT software package that trains a series of multivariate classifiers for each tagging channel via a number of stochastic gradient-boosted decision trees \cite{Keck:2016tk}. The training is performed in a hierarchical manner with final-state particles being reconstructed first from detector information. The decay channels are then built up from these particles as illustrated in Figure \ref{fig:fei}, with the reconstruction of the $B$-mesons performed last. The tagging performance of the FEI exceeds conventional approaches by up to 50$\%$ \cite{Keck:2018lcd}. For each $B$-meson tag ($B_{\mathrm{tag}}$) candidate reconstructed by the FEI, a value of the final multivariate classifier output, the SignalProbability, is assigned. The SignalProbability is distributed between zero and one, representing candidates identified as being background- and signal-like, respectively.

Reduced data samples, or skims, of both data and MC are produced centrally by the Belle II Collaboration, and are available for use in analyses. These include both hadronic and semileptonic skims, and involve the application of the FEI together with a number of loose selections that aim to reduce the sample sizes with little to no loss of signal efficiency.

For the hadronic FEI, the minimum number of tracks per event satisfying certain quality criteria is set to three. The vast majority of $B$-meson decay chains corresponding to the hadronic FEI channels include at least three charged particles, and such a criterion is useful at suppressing background from non-$B\bar{B}$ events. Requirements are placed on the track parameters as defined in Ref.\cite{belle2tracking:2021} to ensure close proximity to the interaction point (IP), with the distance from the center of the detector along the $z$-axis (corresponding to the direction of the electron beam) and in the transverse plane satisfying $|z_0| < 2.0\,\rm cm$ and $|d_0| < 0.5\,\rm cm$, respectively. A minimum threshold $p_t >$ 0.1 GeV/$c$ is placed on the particle transverse momentum. Similar restrictions are applied to the ECL clusters in the event, with at least three clusters required within the polar angle acceptance of the CDC, 0.297 $< \theta <$ 2.618, that satisfy a minimum energy threshold $E > 0.1$ GeV. The total detected energy per event is required to be at least 4 GeV. The total energy deposited in the ECL is restricted to $2\,{\rm GeV} < E_{\rm ECL} < 7\,{\rm GeV}$, however, to suppress events with an excess of energy deposits due to beam background.

The FEI typically results in the order of 20 $B_{\mathrm{tag}}$ candidates per event. The number of these candidates is reduced with selections on the beam-energy-constrained mass, $M_{\mathrm{bc}} = \sqrt{E_{\mathrm{beam}}^{2}/(4c^4)-|\vec{p}^{\,2}_{B_{\mathrm{tag}}}|/c^2}$, and energy difference, $\Delta E = E_{B_{\mathrm{tag}}} - E_{\mathrm{beam}}/2$, where $E_{\mathrm{beam}}$ is the centre-of-mass (CMS) energy of the $e^+e^-$ system, 10.58 GeV, and $\vec{p}_{B_{\mathrm{tag}}}$ and $E_{B_{\mathrm{tag}}}$ are the  $B_{\mathrm{tag}}$ momentum and energy in the CMS frame, respectively. The criteria applied during the hadronic FEI skim are $M_{\mathrm{bc}} > 5.24$ GeV/$c^2$ and $|\Delta E| < 0.2$ GeV.

Finally, a loose requirement on the $B_{\mathrm{tag}}$ classifier output,  SignalProbability $> 0.001$, provides further background rejection with minimal signal loss. 

\section{Event Selection and Analysis Strategy}
\label{sec:selections}

For this analysis, the distribution of the square of the missing mass, $M_{\mathrm{miss}}^2$, is the variable chosen for the determination of the signal yields in data. We define the four-momentum of the signal $B$-meson \Bsig{} in the CMS frame as $p_{B_{\mathrm{sig}}} \equiv  (E_{B_{\mathrm{sig}}}, \vec{p}_{B_{\mathrm{sig}}}) = \left(m_{\Upsilon(4S)}/2, -\vec{p}_{B_{\mathrm{tag}}}\right)$, where $m_{\Upsilon(4S)}$ is the known $\Upsilon$(4S) mass \cite{Zyla:2020zbs}. We set the energy of $B_{\mathrm{sig}}$ to be half of the $\Upsilon$(4S) rest mass, and take the $B_{\mathrm{sig}}$ momentum to be the negative $B_{\mathrm{tag}}$ momentum. We then define the missing four-momentum as $p_{\mathrm{miss}} \equiv (E_{\mathrm{miss}}, \vec{p}_{\mathrm{miss}}) =  p_{B_{\mathrm{sig}}} - p_Y$, where $Y$ represents the combined electron-pion system. The square of the missing momentum can then simply be defined as $M_{\mathrm{miss}}^2 \equiv p_{\mathrm{miss}}^2$.

The event selections applied follow closely those from a 2013 study of exclusive, hadronically tagged $B \to \X_u \ell^+ \nu_\ell$ decays reconstructed in the full 711\invfb Belle data set \cite{Sibidanov:2013sb}. All selections are applied in addition to the hadronic FEI skim criteria detailed in the previous section.

At the event level, a loose selection on the second normalised Fox--Wolfram moment \cite{Wolfram:1978fw} is applied, R$_{2} <$ 0.4, in order to suppress continuum background. To reject incorrectly reconstructed $B_{\mathrm{tag}}$ candidates, the tag-side $M_{\mathrm{bc}}$ criterion is tightened to $M_{\mathrm{bc}} > 5.27$ GeV/c$^2$. The $B_{\mathrm{tag}}$ candidate having the highest value of the SignalProbability classifier output is retained in each event.

For the reconstructed electrons, track impact parameters are used to select tracks originating close to the IP, thereby suppressing background events from beam scattering and radiative effects. Tracks are required to have $z$-axis and transverse-plane distances from the IP of $|dz| < 5\,{\rm cm}$, and $dr < 2$ cm, respectively. 
Only those tracks within the acceptance of the CDC are selected. Electrons are identified through selection criteria on the particle identification variables \cite{Kuhr:B2}. These variables describe the probability that each species of charged particle generates the particle-identification signal observed, and are built from a combination of the information returned from individual sub-detectors. Electron candidates are required to have an identification probability above $0.9$ as assigned by the appropriate reconstruction algorithm. A minimum threshold on the lab-frame momentum is placed on the reconstructed electrons, $p_{\mathrm{lab}} > 0.3$ GeV/$c$. 

The four-momenta of the reconstructed electrons are also corrected in order to account for bremsstrahlung radiation. Electron candidates are separated into three momentum regions, from 0.3 to 0.6 GeV/$c$, from 0.6 to 1.0 GeV/$c$ and above 1.0 GeV/$c$. For the first region, no corrections are applied. For the second region, any energy deposit below 600 MeV in the ECL that is not associated with a track and lies within 4.8$^{\circ}$ of the electron candidate is considered a bremsstrahlung photon. The four-momentum of the photon is added to the electron, and the photon is excluded from the rest of the event. In the third region, the same method is applied, with the angular threshold tightened to 3.4$^{\circ}$ and the photon energy selection relaxed to 1.0 GeV. These thresholds are determined by minimizing the root mean square of the difference between generated and reconstructed electron momenta in simulation. If multiple photons meet this criteria, the one nearest the electron candidate is considered. Finally, a single electron is retained in each event with the highest value of the electron identification probability as described above.

For the reconstructed charged pions, similar impact parameter criteria are applied as those for the electrons, with $dr < 2$ cm and $|dz| < 4$ cm. Similarly, the charged pion tracks are only selected within the CDC acceptance. A selection on the relevant particle identification variable is also applied, with a particle identification probability above $0.6$. The sign of the charge of the reconstructed pion is explicitly required to be opposite that of the electron for the $B^0 \to \pi^- e^+ \nu_e$ case.

In reconstructing neutral pions, different thresholds on the photon-daughter energies are required, depending on the polar direction of the candidate photon. These requirements are $E > 0.080$ GeV for the forward end-cap, $E > 0.030$ GeV for the barrel region and $E > 0.060$ GeV for the backward end-cap. A selection on the diphoton mass is also implemented, with 0.120   $\text{GeV}/c^2  < M(\gamma\gamma) < 0.145$ $\text{GeV}/c^2$. A selection on the cosine of the lab-frame opening angle of the $\pi^0$ photon daughters is also applied in order to reject backgrounds from photon pairs that do not originate from $\pi^0$ decays, cos$\psi_{\gamma\gamma} > 0.25$. 

The CMS frame four-momenta of the reconstructed pion and electron are combined into the system $Y$. The angle between the flight directions of the signal $B$-meson as inferred from initial beam conditions and the $Y$ is then used to select events more likely to originate from the decay of interest. The cosine of this angle, $\mathrm{cos}\theta_{BY}$, is defined as
$$
\mathrm{cos}\theta_{BY} = \frac{2E_{\mathrm{beam}}E_Y - m_{B_{\mathrm{sig}}}^2 - m_Y^2}{2|\vec{p}_{B_{\mathrm{sig}}}||\vec{p}_Y|},
$$
where $m_{B_{\mathrm{sig}}}$ is the invariant mass of the signal $B$-meson, and $E_Y$, $m_Y$ and $\vec{p}_Y$ are the energy, invariant mass and momentum of the $Y$ system, respectively. A value of $|\mathrm{cos}\theta_{BY}| < 1$ is expected if only a neutrino is missing in the reconstruction. However, to account for resolution effects as well as to avoid introducing potential bias in the background $M_{\mathrm{miss}}^2$ distributions, this requirement is loosened to $|\mathrm{cos}\theta_{BY}| < 3$.

To ensure that the reconstructed electron and pion tracks originate from the same vertex in $B^0 \to \pi^- e^+ \nu_e$ decays, the difference between the $z$-coordinates of both tracks at their points of closest approach to the $z$-axis is required to be $|z_e - z_\pi| < 1$ mm.

A minimum threshold on the missing energy, $E_{\mathrm{miss}}$, is placed to account for the neutrino, with $E_{\mathrm{miss}} > 0.3$ GeV. All remaining charged tracks and neutral ECL clusters after the reconstruction of the $\Upsilon$(4S) are combined into a single system known as the rest-of-event. Events in which additional tracks satisfying the conditions $dr < 2$ cm, $|dz| < 5$ cm and $p_t > 0.2$ GeV/$c$ remain after the reconstruction of the $\Upsilon$(4S) are excluded. The sum of the ECL energies in the rest of the event are
required to satisfy the following requirements; $E > 0.10$ GeV, $E > 0.09$ GeV and $E > 0.16$ GeV for the forward end-cap, barrel and backward end-cap regions, respectively. This extra energy is required to be below a maximum value of 
$E_{\mathrm{residual}} < 1.0$ GeV for $B^0 \to \pi^- e^+ \nu_e$, and $E_{\mathrm{residual}} < 0.6$ GeV for $B^+ \to \pi^0 e^+ \nu_e$ candidates. After all analysis selections are applied in the reconstruction of the $B^+ \to \pi^0 e^+ \nu_e$ channel from simulation, approximately 4$\%$ of signal events and 8$\%$ of background events possess multiple $\Upsilon$(4S) candidates. In these cases, a single $\Upsilon$(4S) candidate with the lowest value of $M_{\mathrm{miss}}^2$ is retained per event. Any possible bias introduced with this selection was found to be negligible after performing a dedicated study on simulated MC data. There are no multiple $\Upsilon$(4S) candidates present in the reconstruction of the $B^0 \to \pi^- e^+ \nu_e$ mode.




\section{Results}
\label{sec:results}

\begin{figure}[h!]
\begin{center}
\includegraphics[scale=0.5]{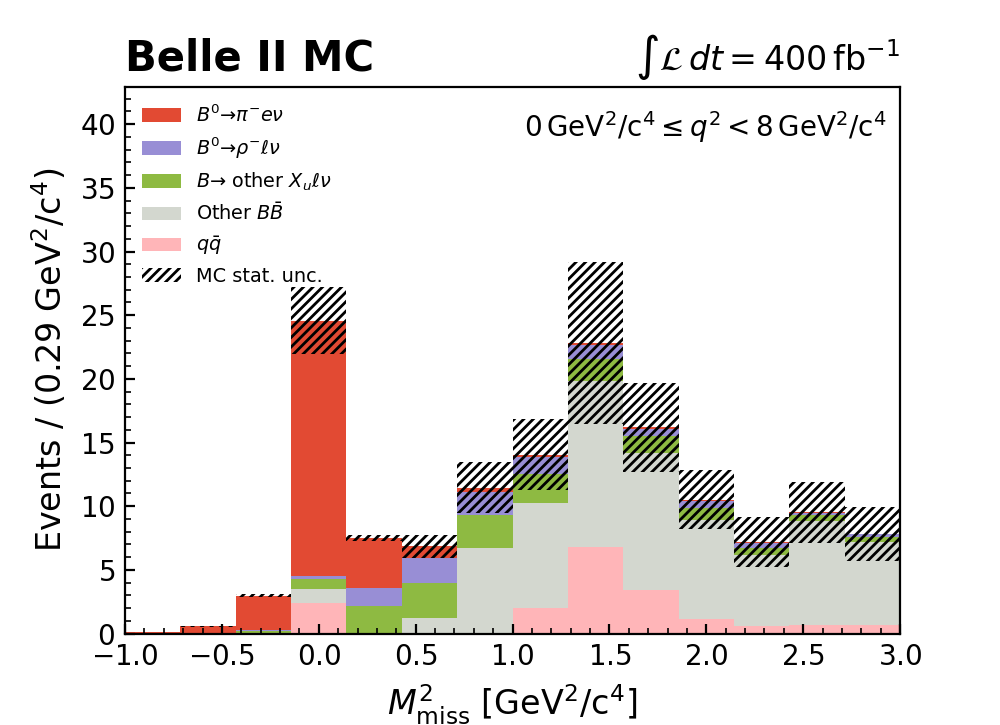}
\includegraphics[scale=0.3]{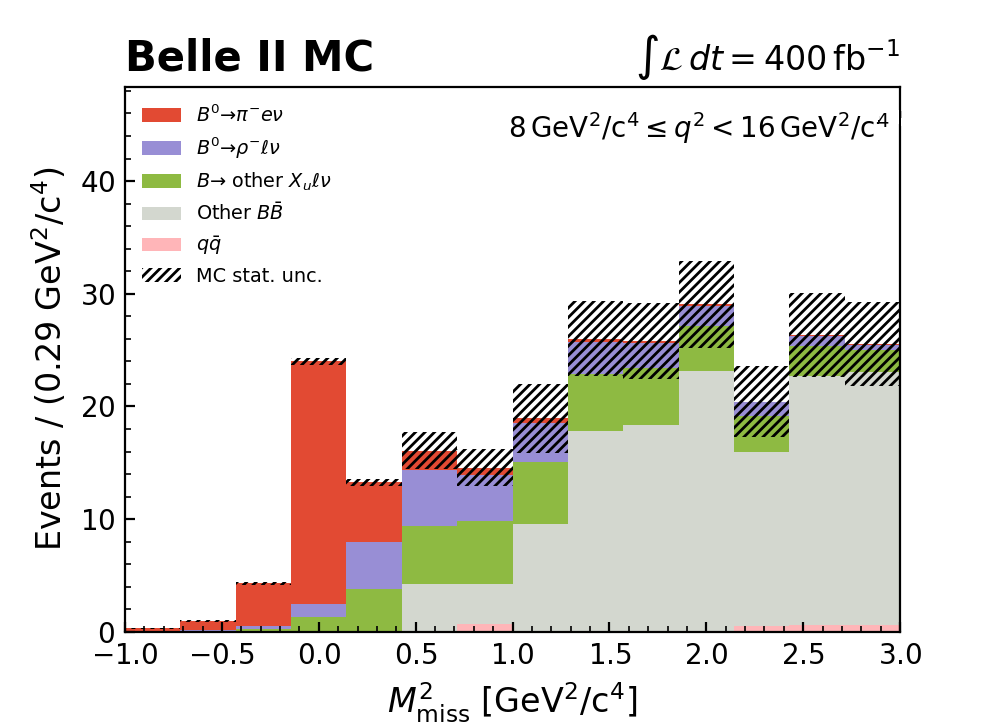}
\includegraphics[scale=0.3]{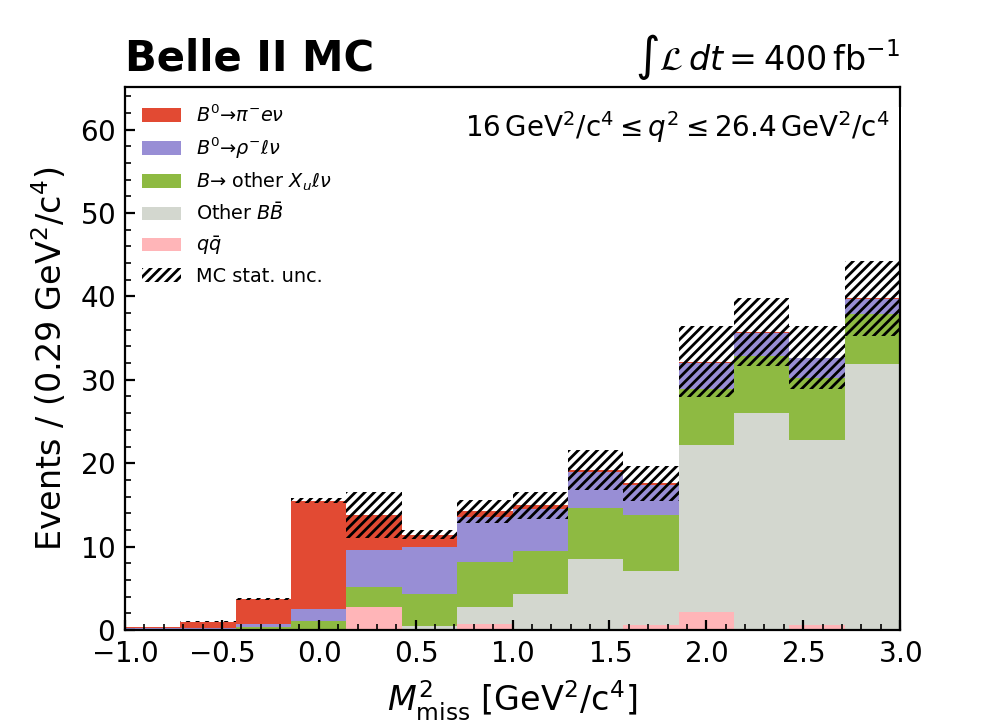}
  \caption{Expected $M_{\mathrm{miss}}^2$ distributions for $B^0 \to \pi^-e^+\nu_e$ candidates restricted to three bins in $q^2$ and reconstructed from a sample corresponding to 400 \invfb of simulated data.}
  \label{fig:prefitpilnuq2B0e}
\end{center}
\end{figure}

\begin{figure}[h!]
\begin{center}
\includegraphics[scale=0.5]{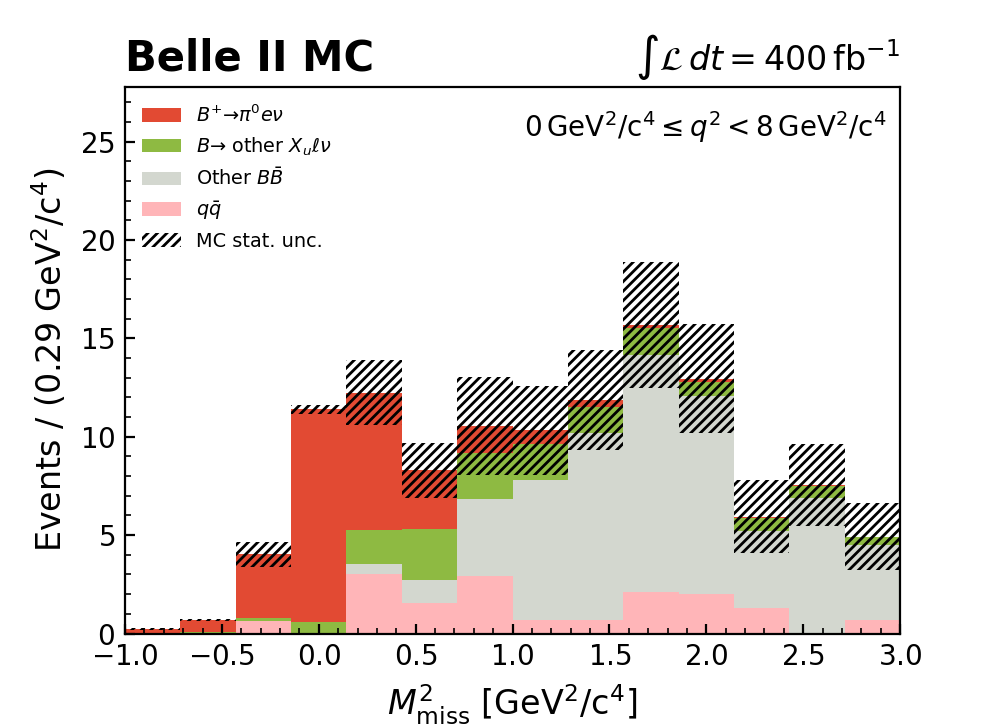}
\includegraphics[scale=0.3]{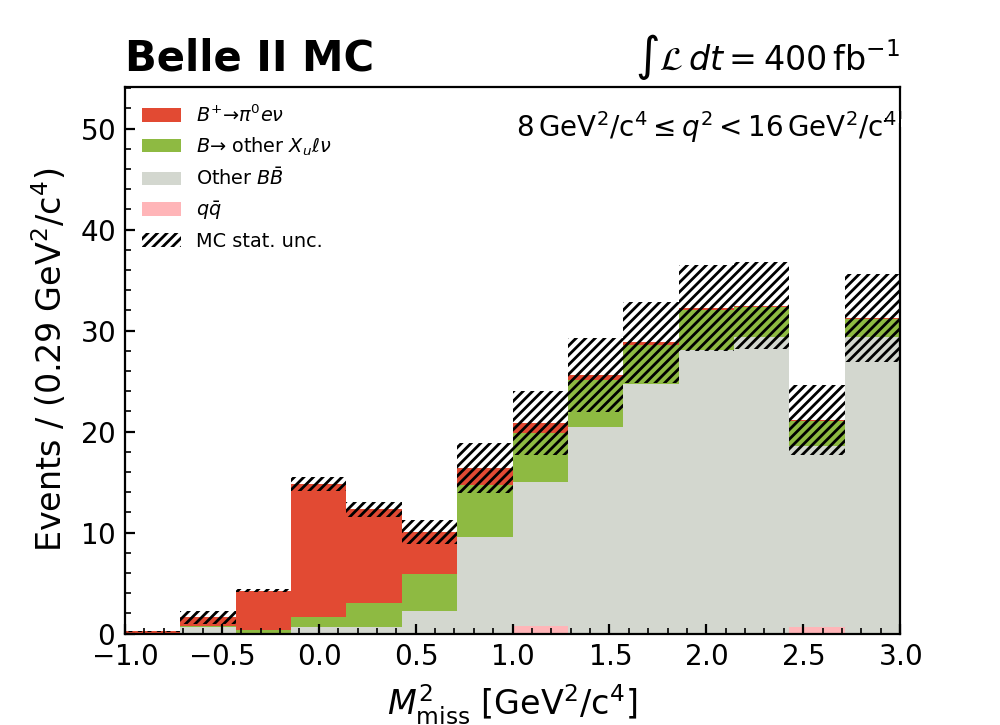}
\includegraphics[scale=0.3]{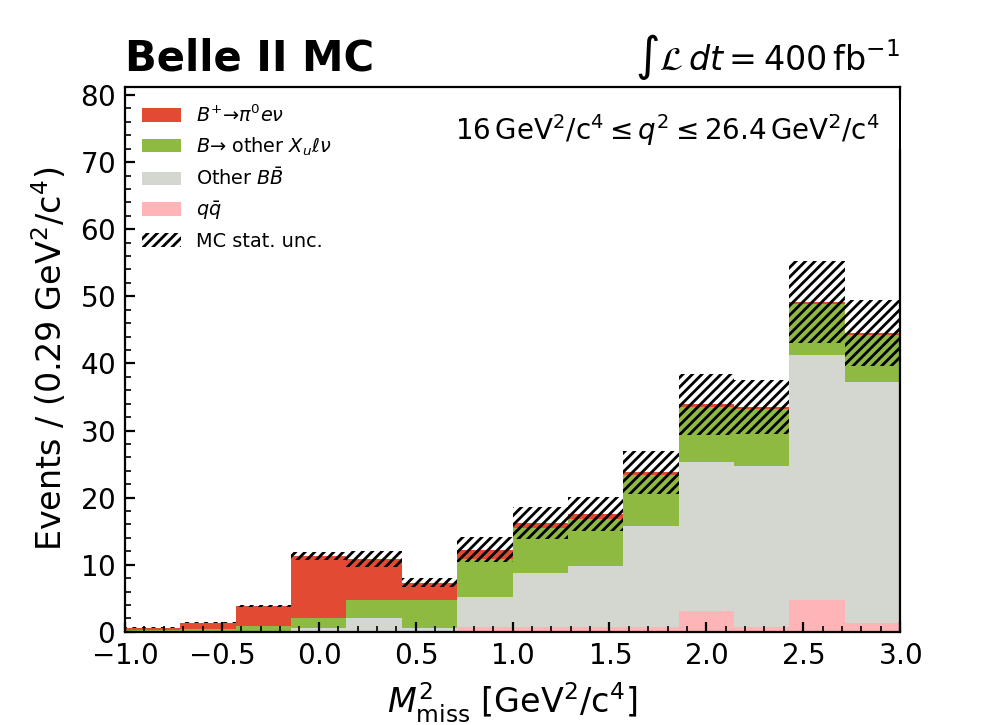}
  \caption{Expected $M_{\mathrm{miss}}^2$ distributions for $B^+ \to \pi^0e^+\nu_e$ candidates restricted to three bins in $q^2$ and reconstructed from a sample corresponding to 400 \invfb of simulated data.}
  \label{fig:prefitpilnuq2Bpe}
\end{center}
\end{figure}

In preparation for the extraction of the CKM-matrix element magnitude $|V_{\mathrm{ub}}|$, the $M_{\mathrm{miss}}^2$ distributions are considered in three bins of the squared momentum transfer to the leptonic system, $q^2$. The resultant $M_{\mathrm{miss}}^2$ distributions in MC after all analysis selections are displayed in Figures \ref{fig:prefitpilnuq2B0e} and \ref{fig:prefitpilnuq2Bpe}, for $B^0\to\pi^-e^+\nu_e$ and $B^+\to\pi^0e^+\nu_e$ decays, respectively. In these distributions, the MC is separated into distinct components to illustrate the relative contributions of various background processes, including the cross-feeds from $B^0 \to \rho^- \ell^+ \nu_\ell$ and other $B \to X_u \ell \nu$ decays, as well as candidates reconstructed from other generic $B\bar{B}$ and continuum events. Due to the small sample size, only three $q^2$ bins are considered at present, with $0 \leq q^2 < 8$ $\text{GeV}^2/c^4 $, $8 \leq q^2 < 16$ $\text{GeV}^2/c^4$, and $16 \leq q^2 \leq 26.4$ $\text{GeV}^2/c^4$, respectively. 


A number of corrections and scaling factors are applied to the simulated data used for Figures \ref{fig:prefitpilnuq2B0e} and \ref{fig:prefitpilnuq2Bpe}. For both reconstructed modes, the total number of MC events is scaled down by a hadronic FEI calibration factor in order to account for the data-MC differences in the tag-side reconstruction efficiency of the FEI. An independent study is performed in order to evaluate these factors for both charged and neutral $B$-meson tags through fitting the electron momentum spectrum in $B\to X e\nu_e$ decays and taking the data-MC ratio of signal events, and the respective values are listed in Table \ref{table:BFparams}. For $B^+ \to \pi^0e^+\nu_e$ decays, an additional scaling factor SF$_{\pi^0} = 1.017 \pm 0.044$ is also applied to the total MC component to correct for data-MC differences in the $\pi^0$ reconstruction efficiency. This factor is also determined via an independent study of $D^0\to K^-\pi^+\pi^0$ and $D^0\to K^-\pi^+$ decays, in which the ratio of signal events in data and MC is determined by fitting the invariant mass of the reconstructed $D^0$ meson.

Furthermore, each MC component is weighted by a set of corrections to account for the differences in the electron identification efficiencies and the pion and kaon misidentification rates between MC and data. These corrections are obtained in an independent study \cite{LeptonID:2318} and are evaluated per event based on the magnitude of the lab-frame momentum $p$ and polar angle $\theta$ of the reconstructed electron tracks. For the $B^0 \to \pi^-e^+\nu_e$ decays, a similar set of MC corrections are applied for the charged pion identification efficiencies and the misidentification rates due to charged kaons.

The event selection criteria are applied to data along with a data-specific correction factor, whereby charged particle momenta are multiplied by a factor of 0.99976 $\pm$ 0.00055 to correct for momentum-scale differences between data and MC.

\begin{figure}[h!]
\begin{center}
\includegraphics[scale=0.3]{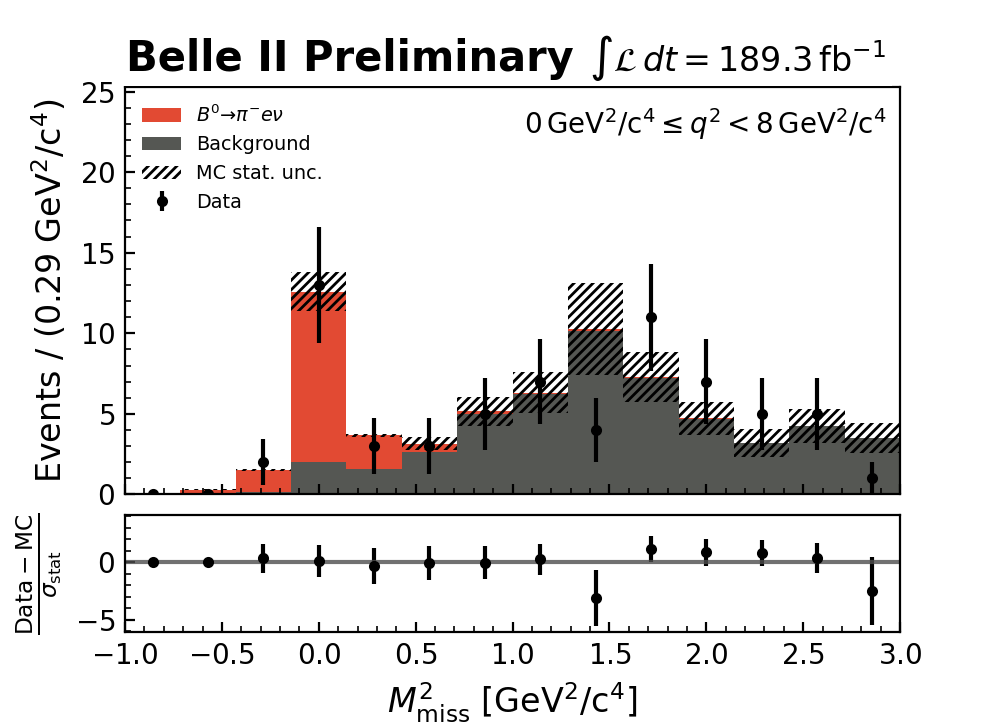}
\includegraphics[scale=0.3]{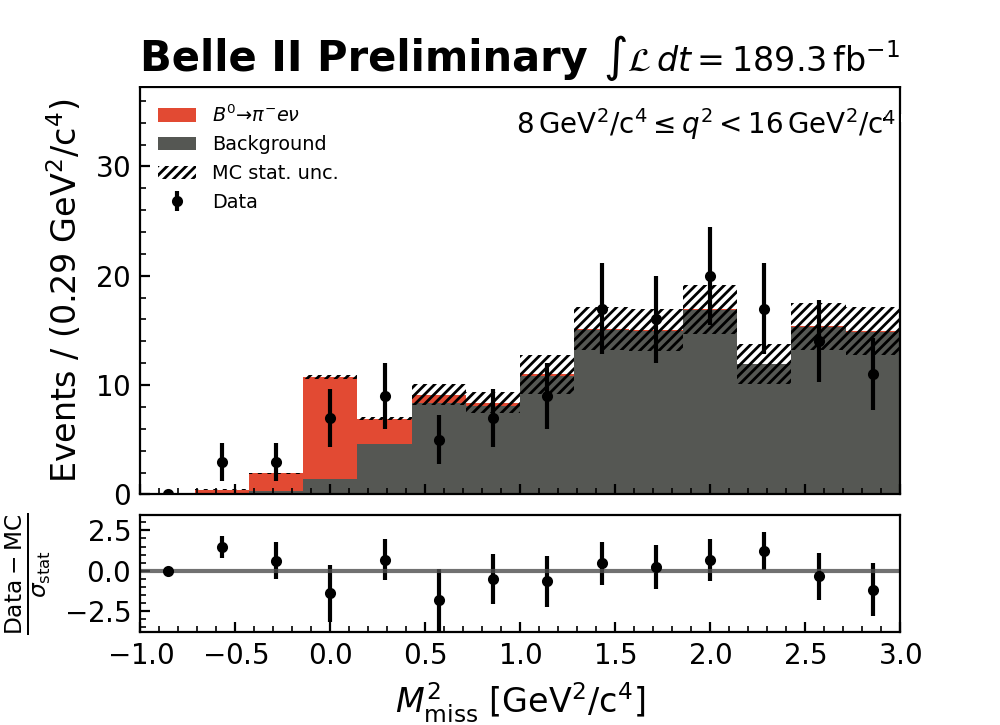}
\includegraphics[scale=0.3]{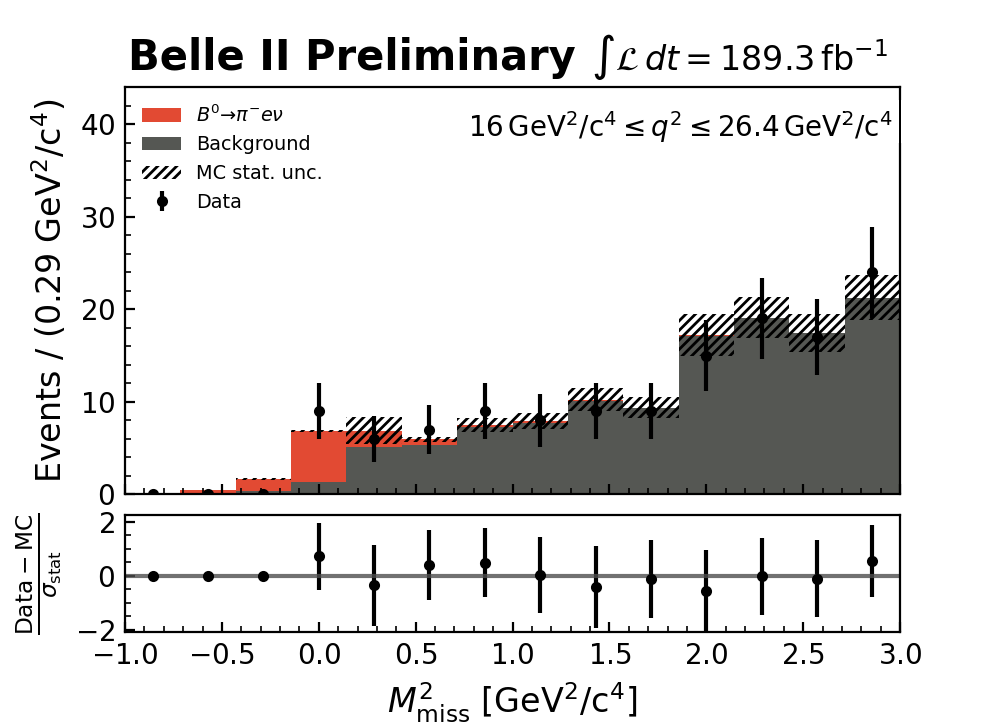}
  \caption{$M_{\mathrm{miss}}^2$ for $B^0 \to \pi^-e^+\nu_e$ decays candidates restricted to three bins in $q^2$ and reconstructed from a sample corresponding to 189.3 \invfb of data with fit projections overlaid.}
  \label{fig:postfitpilnuq2B0e}
\end{center}
\end{figure}

\begin{figure}[h!]
\begin{center}
\includegraphics[scale=0.3]{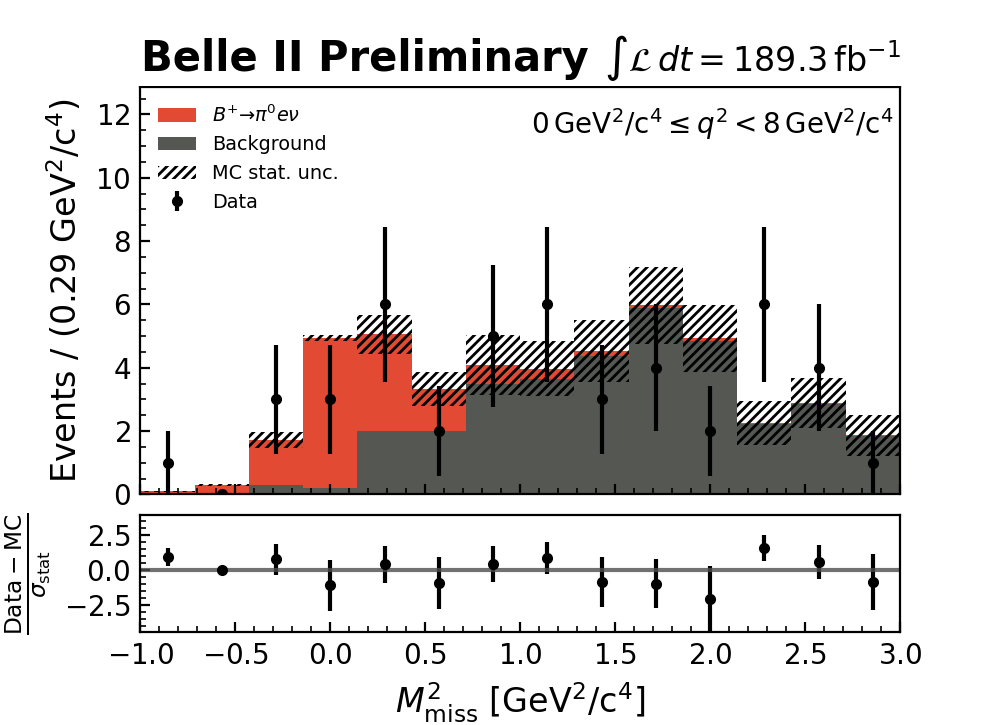}
\includegraphics[scale=0.3]{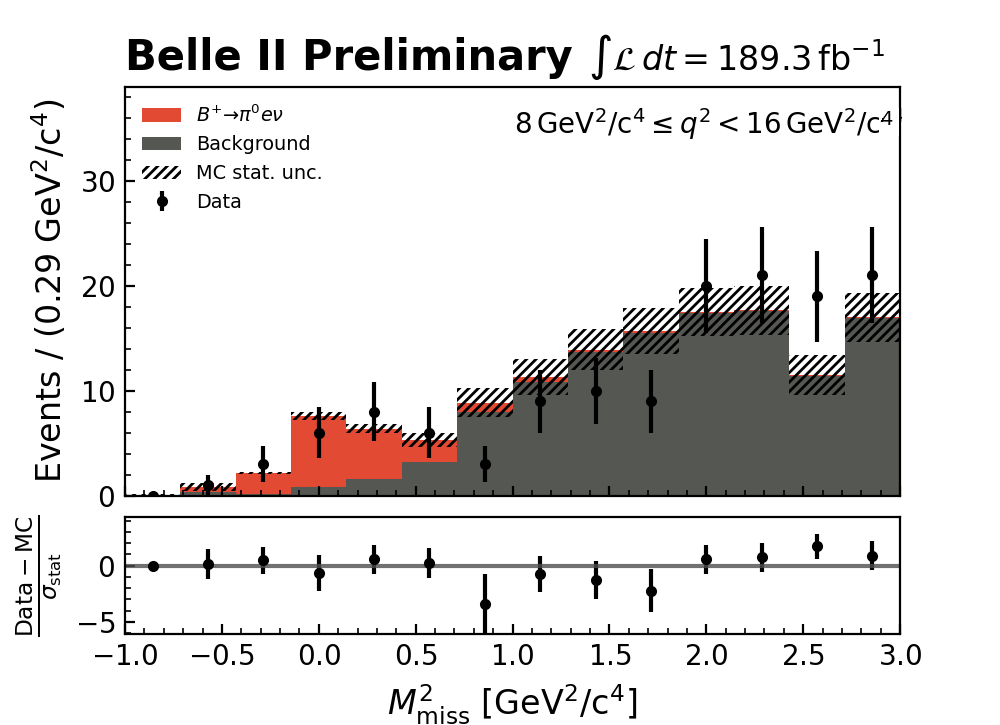}
\includegraphics[scale=0.3]{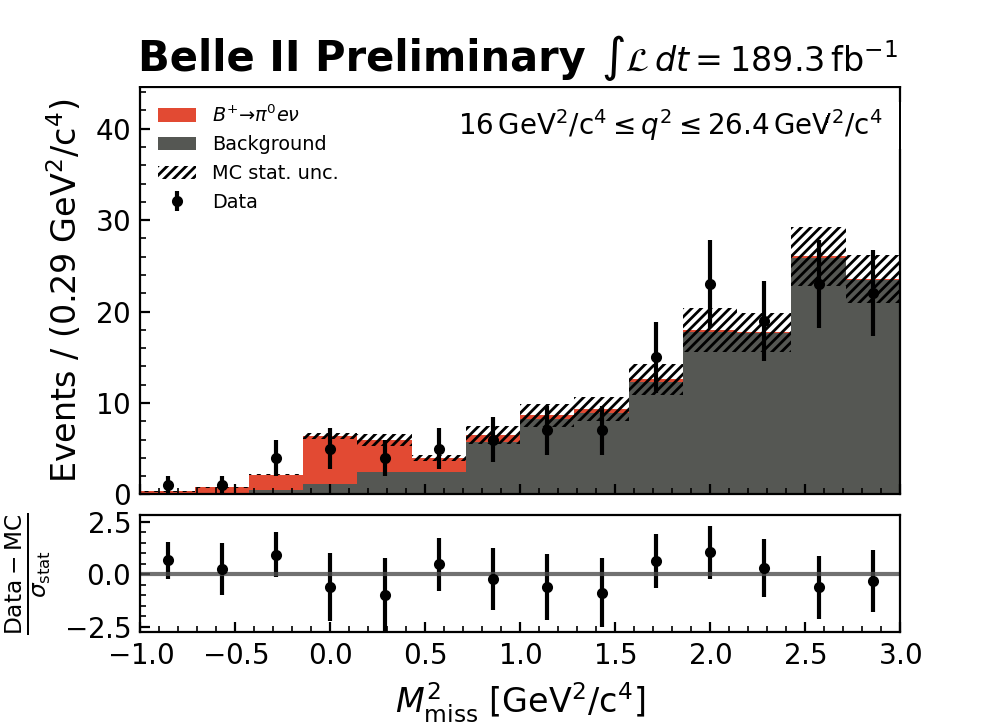}
  \caption{$M_{\mathrm{miss}}^2$ for $B^+ \to \pi^0e^+\nu_e$ decays candidates restricted to three bins in $q^2$ and reconstructed from a sample corresponding to 189.3 \invfb of data with fit projections overlaid.}
  \label{fig:postfitpilnuq2Bpe}
\end{center}
\end{figure}

Template probability density functions (PDFs) are subsequently built from the MC signal and background $M_{\mathrm{miss}}^2$ distributions shown in Figures \ref{fig:prefitpilnuq2B0e} and \ref{fig:prefitpilnuq2Bpe}, normalized to the luminosity of the data sample. For each decay mode studied, due to limited sample size, all background components are combined together into a single background PDF, and a two-component extended unbinned maximum-likelihood fit to the $M_{\mathrm{miss}}^2$ distributions in data is performed. The fit returns two parameters, namely the signal and background yields, which are allowed to float during the fit with no additional constraints. The resultant fitted distributions are shown in Figures \ref{fig:postfitpilnuq2B0e} and \ref{fig:postfitpilnuq2Bpe}, for $B^0\to\pi^-e^+\nu_e$ and $B^+\to\pi^0e^+\nu_e$ decays, respectively. Fairly good agreement between simulated and measured data is observed across the $M_{\mathrm{miss}}^2$ distributions shown, including in the signal region. A clear signal peak can be seen at $M_{\mathrm{miss}}^2$ near zero for both data and MC, with all other backgrounds peaking at higher values of $M_{\mathrm{miss}}^2$.

Additional unbinned maximum-likelihood fits are then performed to the same data samples under the background-only hypothesis. The likelihood ratio $\lambda$ between both fits is computed for each decay mode, $\lambda = \mathcal{L}_{S+B}/\mathcal{L}_{B} \hspace{0.5em},$ where $\mathcal{L}_{S+B}$ and $\mathcal{L}_{B}$ are the maximised likelihoods returned by the fits to the signal-plus-background and background-only hypotheses, respectively. A significance estimator $\Sigma$ is subsequently defined based on the likelihood ratio, $\Sigma = \sqrt{2\ln\lambda}\hspace{0.5em}$. The fitted yields for each decay mode are listed in Table \ref{table:datafitresults}, together with the observed significances.

\begin{table}
\caption{Yields obtained from the maximum-likelihood fits to 189.3\invfb of data. The observed significances are also listed.}\label{table:datafitresults}
\def\arraystretch{1.2}%
\begin{tabular}{cp{10em}p{13em}p{11em}}
\hline\hline
$q^2$ bin & Fitted signal yield & Fitted background yield & Observed significance \\
\hline
\multicolumn{4}{c}{$B^0 \to \pi^- e^+ \nu_e$}\\\hline
0 $\leq q^2 < 8$ GeV$^2$  & 15.3 $\pm$ 4.5 & 50.7 $\pm$ 7.5 & 5.4$\sigma$ \\
8 $ \leq q^2 < 16$ GeV$^2$  & 15.3 $\pm$ 4.8 & 123 $\pm$ 11 & 5.3$\sigma$ \\
16 $ \leq q^2 \leq 26.4$ GeV$^2$  & 10.4 $\pm$ 4.2 & 122 $\pm$ 11 & 3.8$\sigma$ \\
Sum & 41.0 $\pm$ 7.8 & 295 $\pm$ 18 & --\\\hline
\multicolumn{4}{c}{$B^+ \to \pi^0 e^+ \nu_e$}\\\hline
0 $ \leq q^2 < 8$ GeV$^2$  & 12.2 $\pm$ 4.4 & 33.8 $\pm$ 6.4 & 4.6$\sigma$ \\
8 $ \leq q^2 < 16$ GeV$^2$  & 18.3 $\pm$ 5.1 & 118 $\pm$ 11 & 5.8$\sigma$ \\
16 $ \leq q^2 \leq 26.4$ GeV$^2$  & 15.2 $\pm$ 5.1 & 127 $\pm$ 12 & 4.4$\sigma$ \\
Sum & 45.6 $\pm$ 8.5 & 278 $\pm$ 18 & --\\\hline
\hline\hline
\end{tabular}
\end{table}

Before utilising the signal yields obtained from the fitted $M_{\mathrm{miss}}^2$ distributions in the determination of the $B \to \pi e^+\nu_e$ branching fractions, an unfolding procedure is implemented in order to correct these signal yields. When data or MC is reconstructed by the Belle II detector, due to resolution effects, the true underlying distributions of variables are smeared, resulting in reconstructed distributions that are somewhat offset from truth. Unfolding refers to a process that aims to recover these true distributions. In this analysis, we unfold the distribution of signal yields obtained from the $M_{\mathrm{miss}}^2$ fits in multiple $q^2$ bins.

\begin{figure}[h!]
\begin{center}
\includegraphics[scale=0.09]{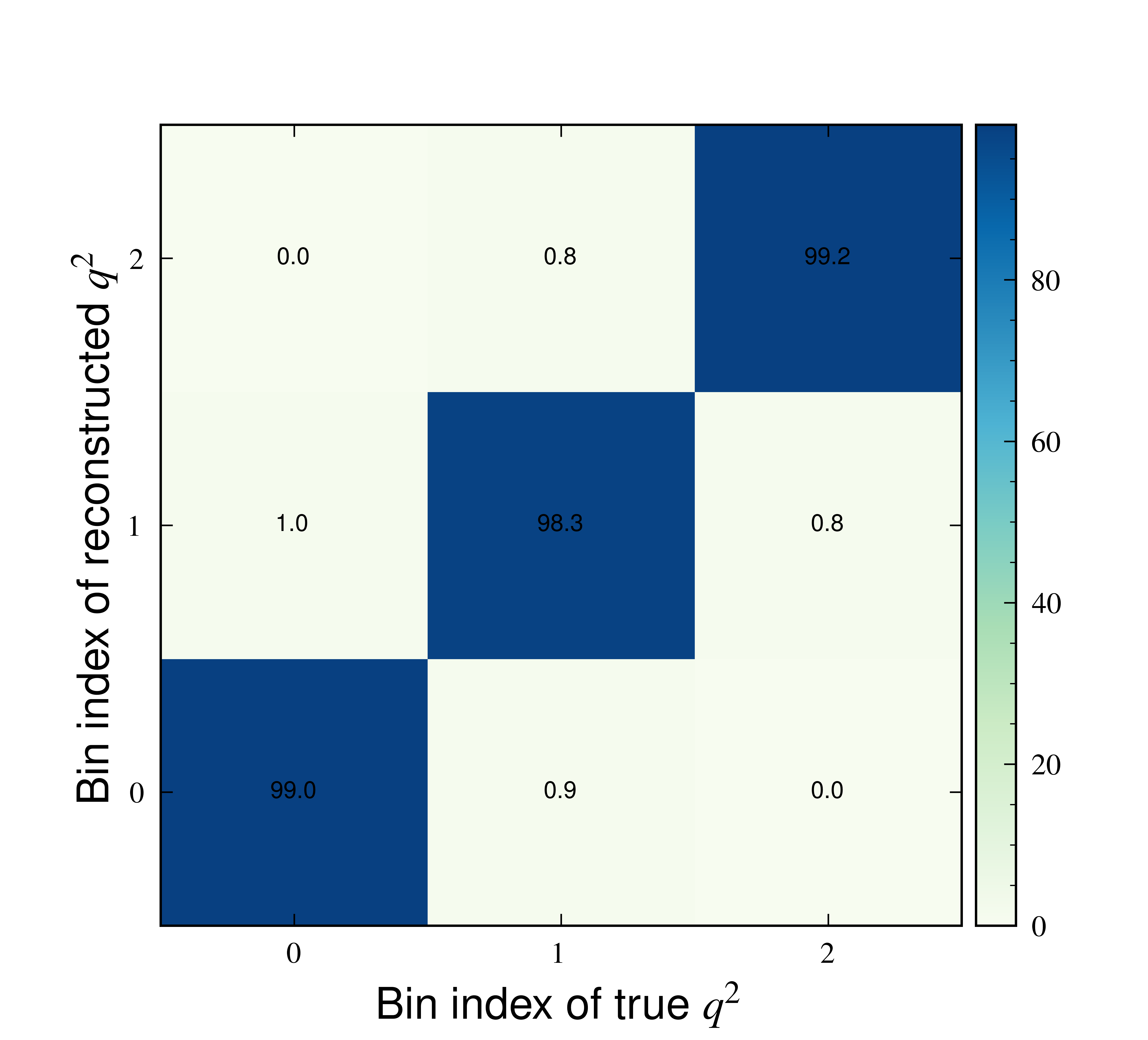}
\includegraphics[scale=0.55]{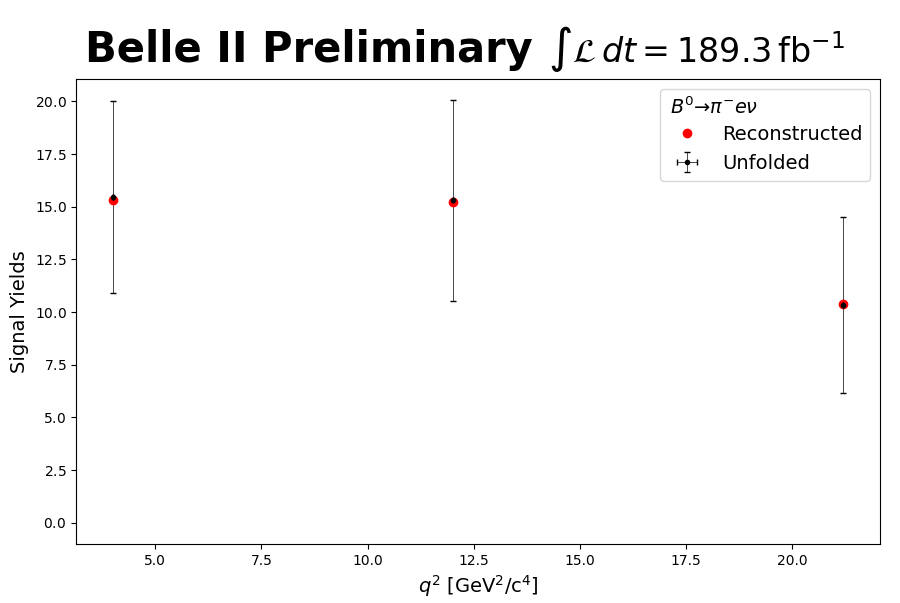}
  \caption{The $q^2$ bin migration matrix in percent (top) and comparison of the reconstructed and unfolded signal yields (bottom) in 189.3\invfb of data, for $B^0\to\pi^- e^+\nu_e$.}
  \label{fig:unfoldingB0}
\end{center}
\end{figure}

\begin{figure}[h!]
\begin{center}
\includegraphics[scale=0.09]{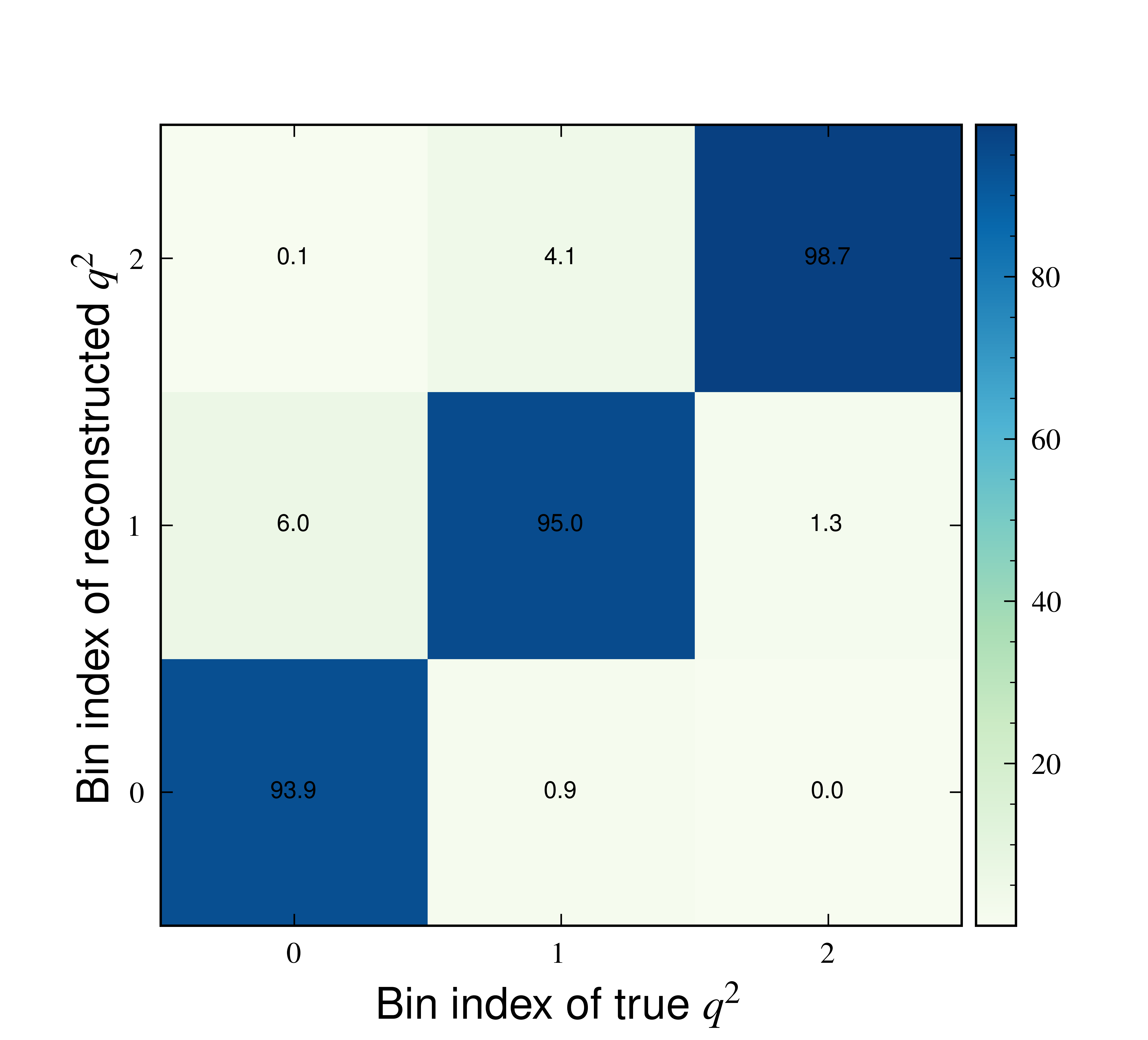}
\includegraphics[scale=0.55]{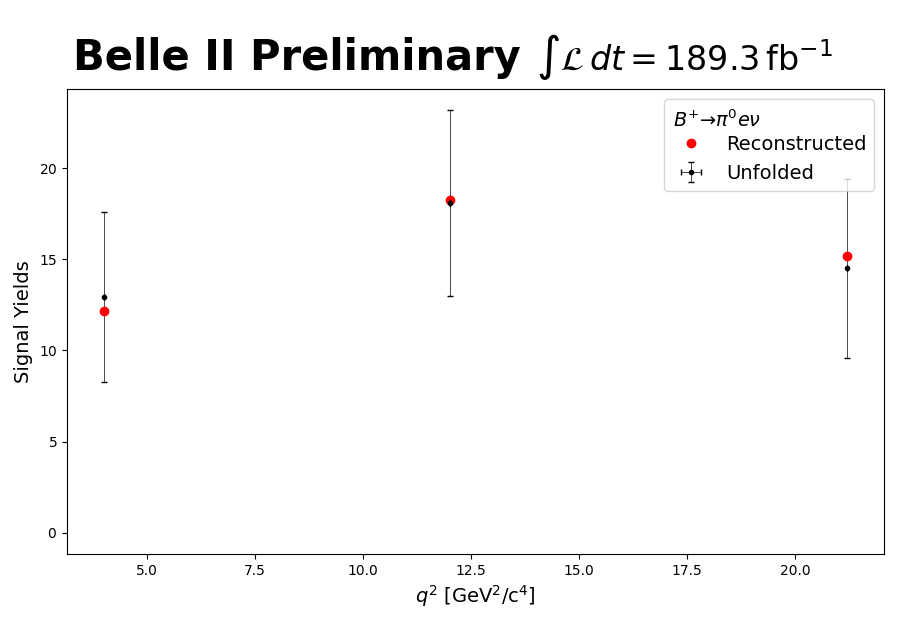}
  \caption{The $q^2$ bin migration matrix in percent (top) and comparison of the reconstructed and unfolded signal yields (bottom) in 189.3\invfb of data, for $B^+\to\pi^0 e^+\nu_e$.}
  \label{fig:unfoldingBp}
\end{center}
\end{figure}

Using MC, for each decay mode, we access both the number of reconstructed and true events in each $q^2$ bin. These are then used to define a bin migration matrix $M_{ij}$, which lists the proportions of events in each reconstructed bin $i$ for each true bin $j$. The migration matrices for $B^0\to\pi^-e^+\nu_e$ and $B^+\to\pi^0e^+\nu_e$ are illustrated in Figures \ref{fig:unfoldingB0} and \ref{fig:unfoldingBp}, respectively. Due to the high purity of the hadronically tagged approach and the fact that the current number of $q^2$ bins considered is quite small, there is very little bin migration, and the vast majority of events are reconstructed within the same $q^2$ bin as the underlying MC truth in each case. A simple bin-by-bin correction is thus used for the $q^2$ unfolding, provided by the \texttt{RooUnfold} package \cite{cern:roounfold}. The reconstructed and unfolded yields are plotted for each $q^2$ bin, as shown in Figures \ref{fig:unfoldingB0} and \ref{fig:unfoldingBp}. At the current data sample size, the effect of the unfolding is minimal, and produces a slight shift in the signal yields well within the statistical errors.  

Given the unfolded fitted yields obtained from data in each $q^2$ bin, the partial branching fractions $\mathcal{B}_i$ for the $B\to\pi e^+\nu_e$ decays were then extracted using the following formulae:\\

\begin{equation}
\Delta \mathcal{B}_i(B^0 \to \pi^- e^+ \nu_e) = \frac{N_{\mathrm{sig, i}}^{\mathrm{data}}(1 + f_{\mathrm{+0}})}{2\times \mathrm{CF}_{\mathrm{FEI}} \times N_{B\bar{B}} \times \epsilon_i} \hspace{0.5em},
\end{equation}
\begin{equation}
\Delta \mathcal{B}_i(B^+ \to \pi^0 e^+ \nu_e) = \frac{N_{\mathrm{sig, i}}^{\mathrm{data}}(1 + f_{\mathrm{+0}})}{2\times \mathrm{CF}_{\mathrm{FEI}} \times N_{B\bar{B}} \times  \mathrm{SF}_{\pi^0} \times f_{\mathrm{+0}} \times \epsilon_i} \hspace{0.5em},
\end{equation}
where $N_{\mathrm{sig, i}}^{\mathrm{data}}$ is the unfolded fitted signal yield obtained from data in each $q^2$ bin $i$, $f_{\mathrm{+0}}$ is the ratio between the branching fractions of the decays of the $\Upsilon$(4S) meson to pairs of charged and neutral $B$-mesons \cite{Zyla:2020zbs}, $\mathrm{CF}_{\mathrm{FEI}}$ is the FEI calibration factor, $\mathrm{SF}_{\pi^0}$ is a scaling factor to correct the $\pi^0$ reconstruction efficiency (for $B^+ \to \pi^0 e^+ \nu_e$ only), $N_{B\bar{B}}$ is the number of $B$-meson pairs counted in the current data set, and $\epsilon_i$ is the signal reconstruction efficiency in each $q^2$ bin. The factor of two present in the denominator accounts for the two $B$-mesons in the $\Upsilon$(4S) decay. The signal efficiency is calculated from the ratio of signal events present in the MC sample before and after all analysis selections. Table \ref{table:BFparams} lists the values of the above parameters used as input to the measurements, and the resultant partial branching fractions are listed in Tables {\ref{table:BFpiB0}} and {\ref{table:BFpiBp}} for $B^0 \to \pi^- e^+ \nu_e$ and $B^+ \to \pi^0 e^+ \nu_e$ decays, respectively. Each set of partial branching fractions is summed to provide the total branching fractions, and these are listed together with the branching fractions obtained by fitting the $M_{\mathrm{miss}}^2$ distributions over the full $q^2$ range.  The branching fractions agree with the current world averages \cite{Zyla:2020zbs} in all cases. All branching fraction uncertainties are largely dominated by sample size at the current integrated luminosity.

\begin{table} [h!]
\caption{Values of the parameters used in the extraction of the $B\to\pi e^+\nu_e$ partial branching fractions.}\label{table:BFparams}
\def\arraystretch{1.2}%
\begin{tabular}{p{8em}p{8em}p{8em}}
\hline\hline
  Parameter                                  & $B^0 \to \pi^- e^+ \nu_e$   & $B^+ \to \pi^0 e^+ \nu_e$\\
                                    \hline
 $f_{\mathrm{+0}}$\cite{Zyla:2020zbs}                  & \multicolumn{2}{c}{1.058 $\pm$ 0.024}\\  
  $N_{B\bar{B}}$                     & \multicolumn{2}{c}{(197.2 $\pm$ 5.7) $\times 10^6$} \\ 
 $\mathrm{SF}_{\pi^0}$              & --                                                   &   1.030 $\pm $ 0.049\\  
 $\mathrm{CF}_{\mathrm{FEI}}$       &  0.70 $\pm$ 0.02                                & $0.65 \pm 0.02$ \\ 
 \hline\hline
\end{tabular}
\end{table}

\begin{figure} [h!]
\includegraphics[width=0.32\textwidth]{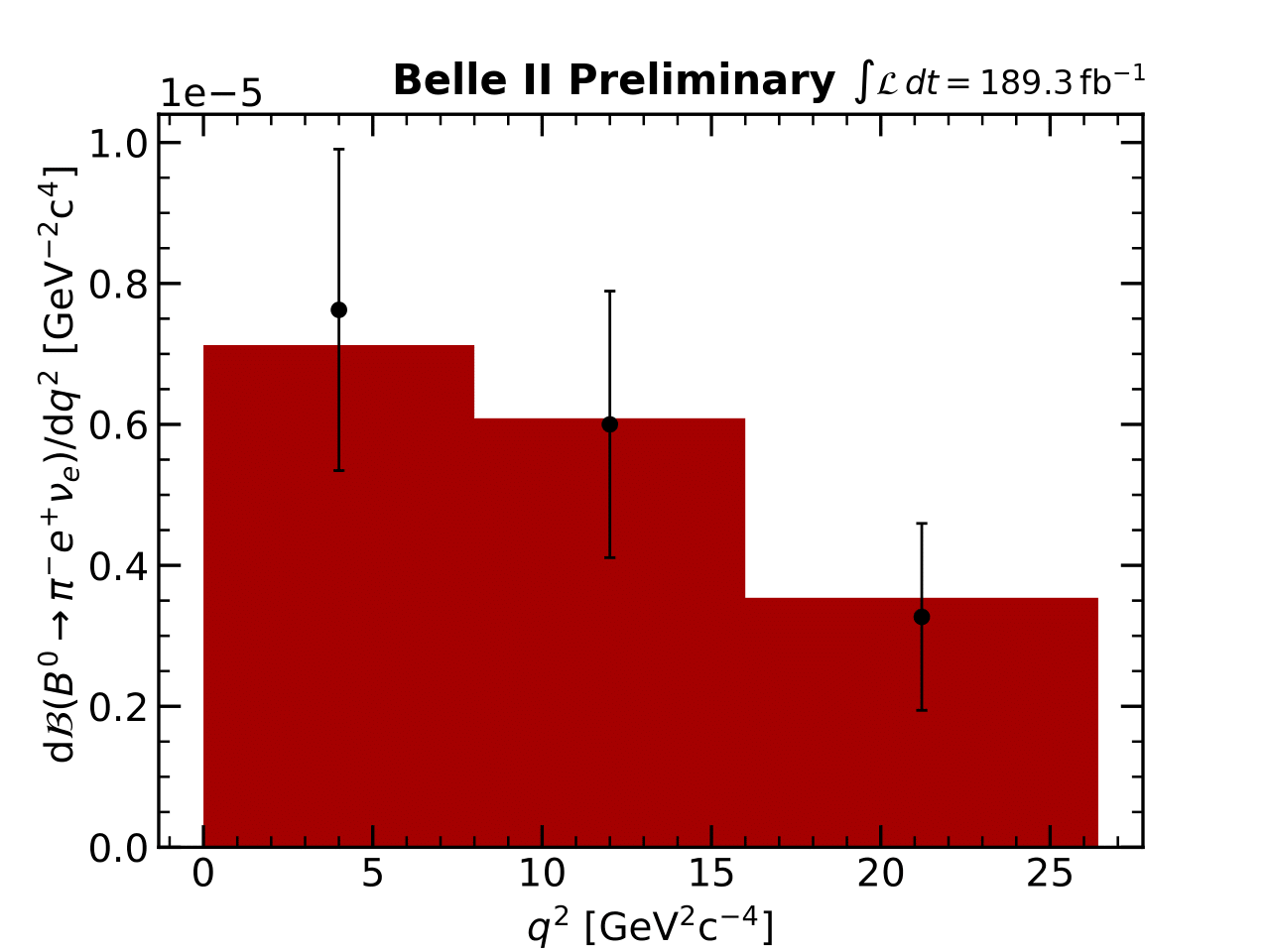}
\includegraphics[width=0.32\textwidth]{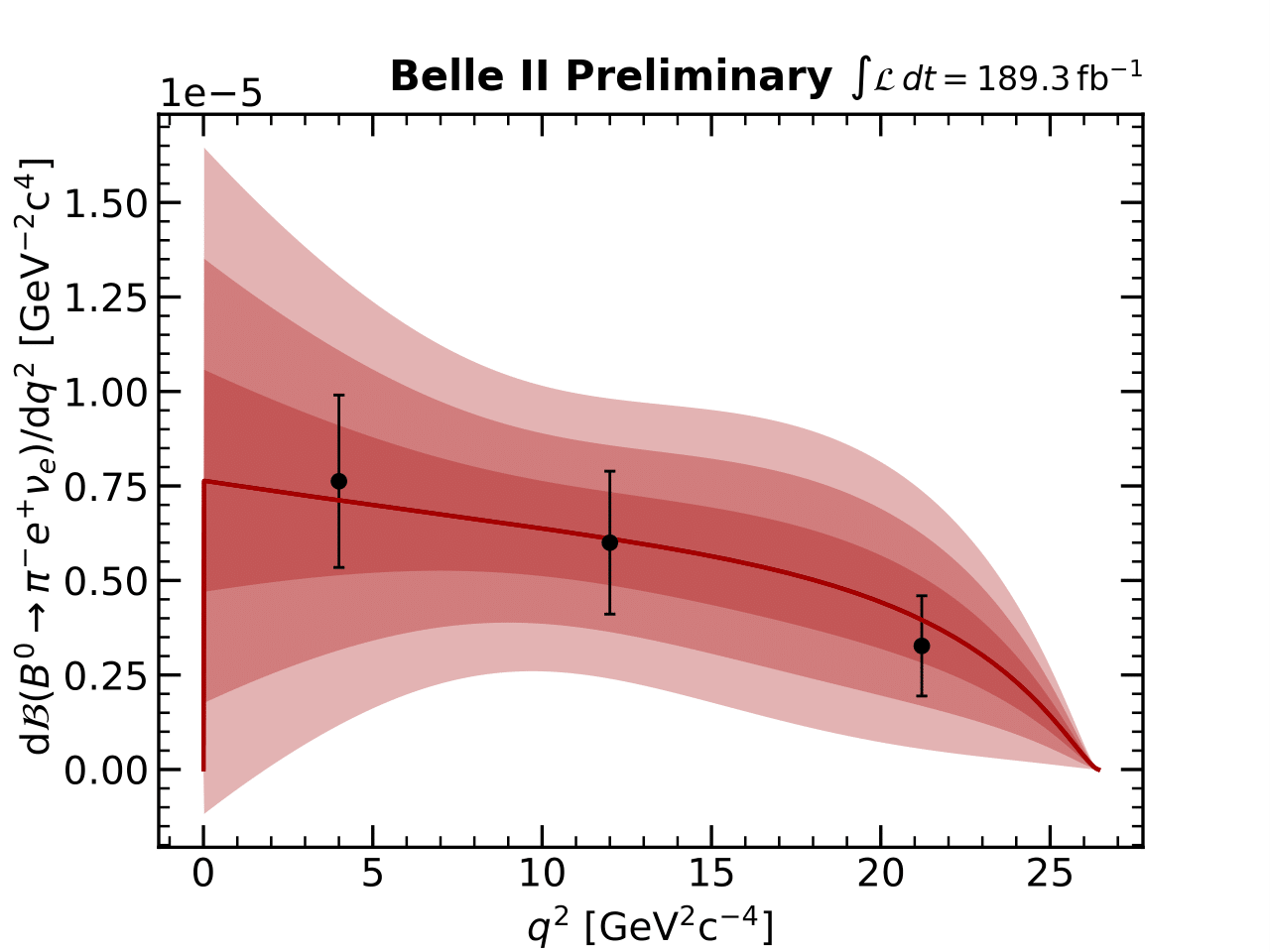}
  \caption{Partial branching fractions of $B^0\to\pi^-e^+\nu_e$ as a function of $q^2$ with fit projections overlaid (left), and with 1-3$\sigma$ uncertainty bands (right),  from 189.3\invfb of data.}
  \label{fig:vubB0e}
\end{figure}

\begin{figure} [h!]
\includegraphics[width=0.32\textwidth]{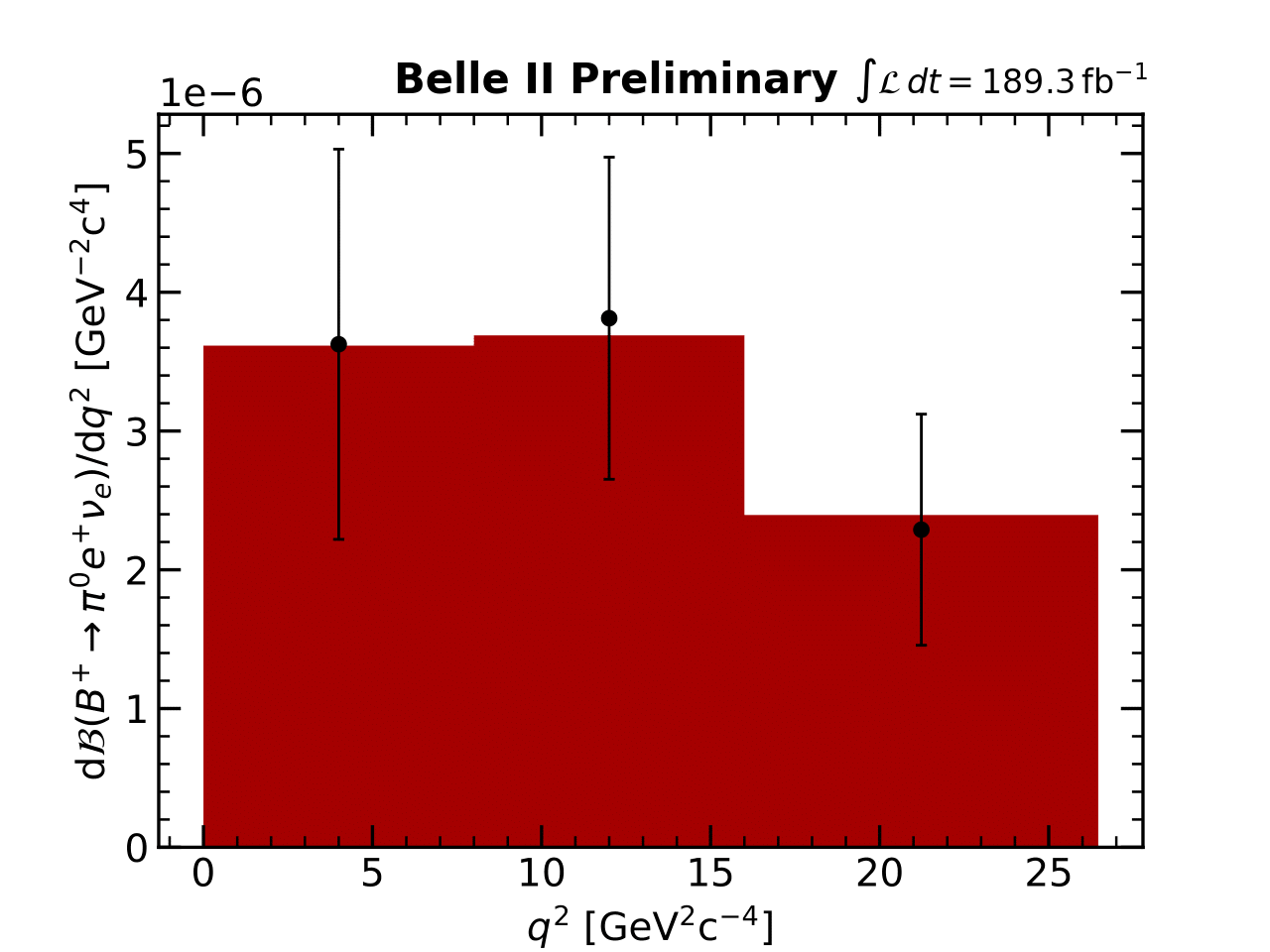}
\includegraphics[width=0.32\textwidth]{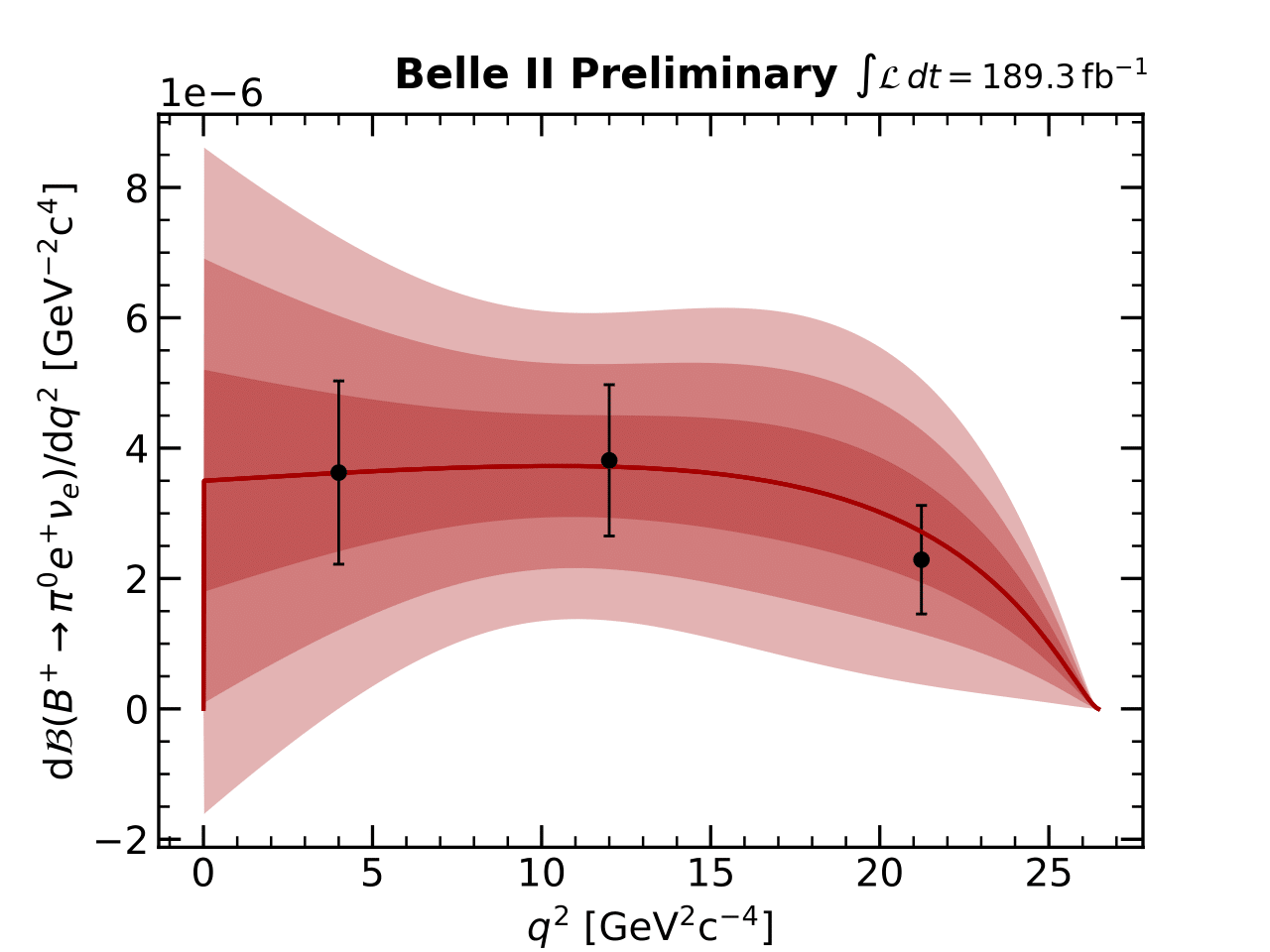}
  \caption{Partial branching fractions of $B^+\to\pi^0e^+\nu_e$ as a function of $q^2$ with fit projections overlaid (left), and with 1-3$\sigma$ uncertainty bands (right),  from 189.3\invfb of data.}
  \label{fig:vubBpe}
\end{figure}

\begin{figure} [h!]
\includegraphics[width=0.32\textwidth]{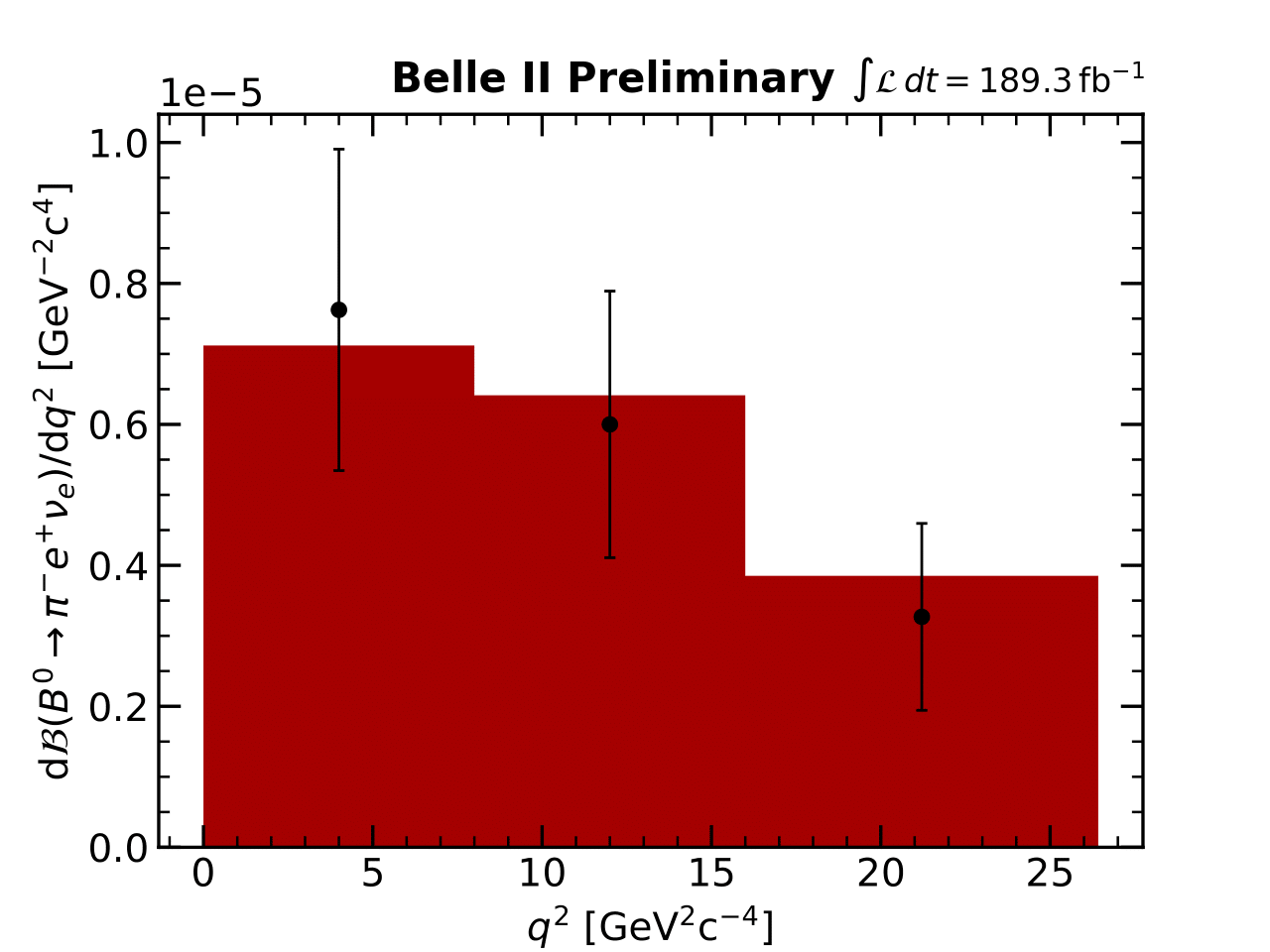}
\includegraphics[width=0.32\textwidth]{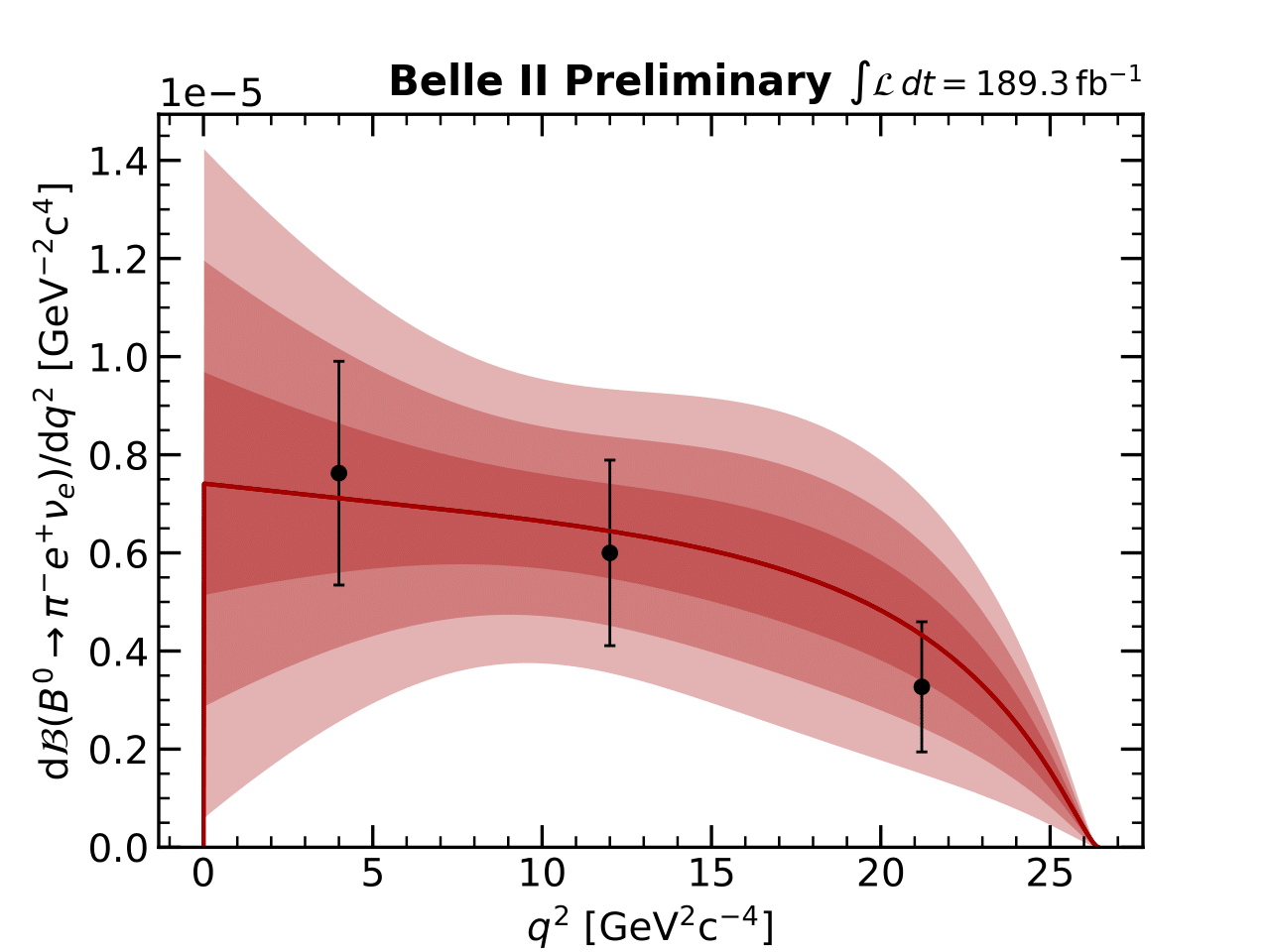}\\
\includegraphics[width=0.32\textwidth]{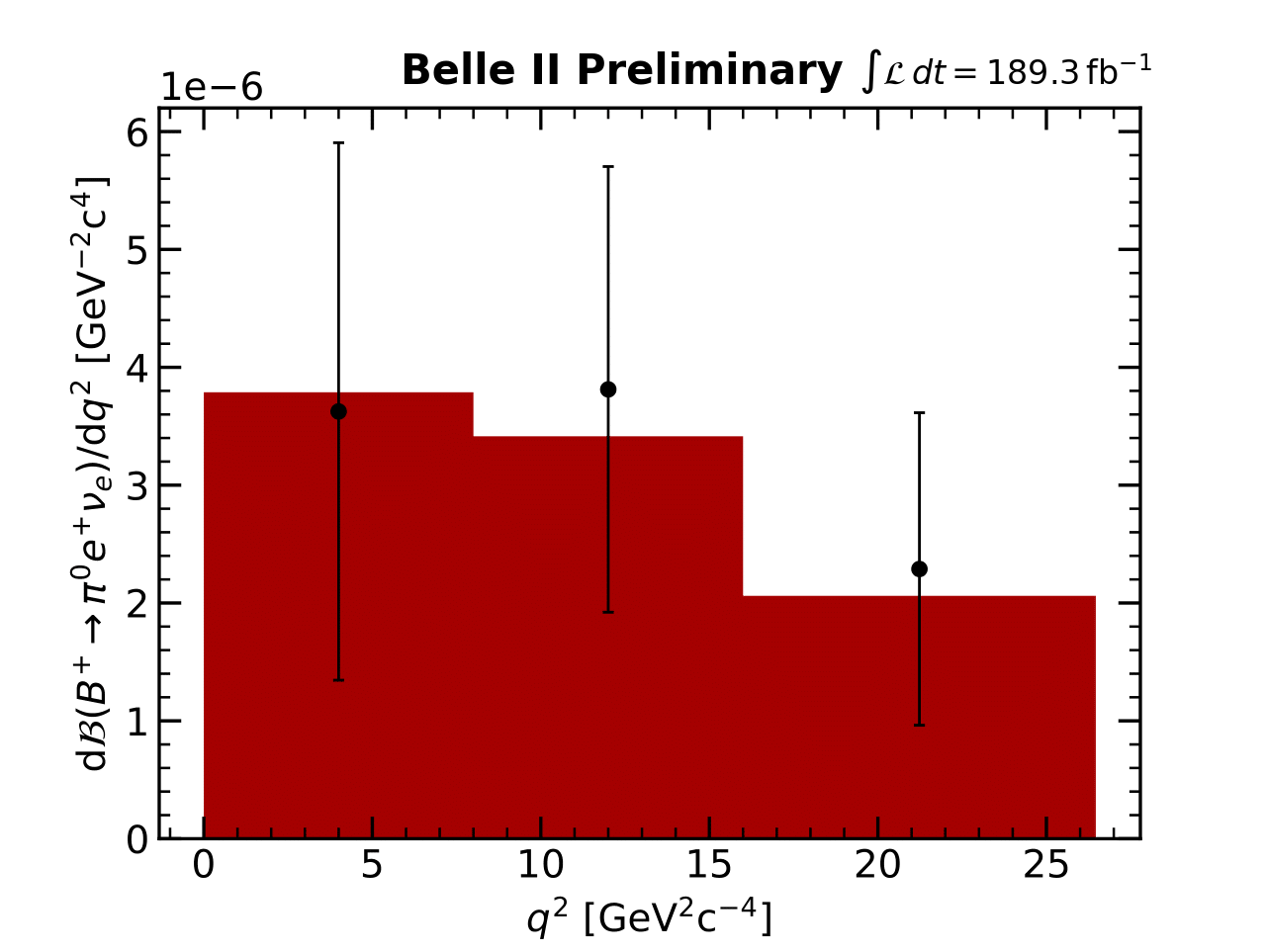}
\includegraphics[width=0.32\textwidth]{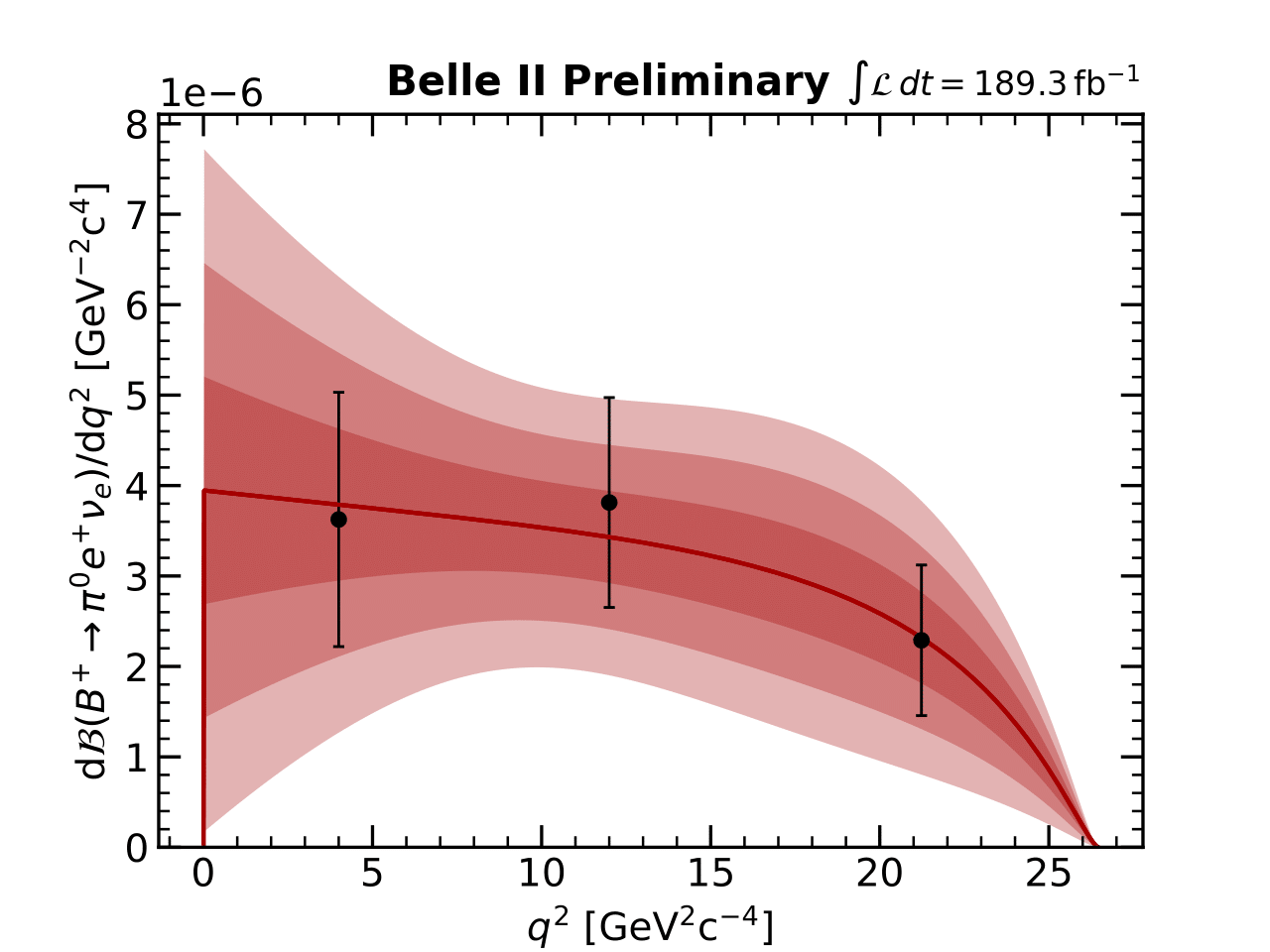}
  \caption{Partial branching fractions of $B^0\to\pi^-e^+\nu_e$ and $B^+\to\pi^0e^+\nu_e$ as a function of $q^2$ with fit projections overlaid (left), and with 1-3$\sigma$ uncertainty bands (right),  from 189.3\invfb of data.}
  \label{fig:vubcombined}
\end{figure}

Finally, first measurements of the magnitude of the CKM-matrix element $|V_{\mathrm{ub}}|$ are extracted from the $B^0\to\pi^-e^+\nu_e$ and $B^+\to\pi^0e^+\nu_e$ analyses, for the current integrated luminosity of 189.3\invfb. For each decay mode, a simultaneous $\chi^2$ fit is performed to the Fermilab/MILC constraints from lattice quantum chromodynamics given in Ref.\cite{Bailey:2015bl} and the measured partial branching fractions from the given data sample. A BCL parameterisation is used for the form factors in the fit \cite{BCL}, with the input and post-fit parameters as detailed in Appendix B. An additional combined simultaneous $\chi^2$ fit is then performed to both partial branching fraction distributions derived from $B^0\to\pi^-e^+\nu_e$ and $B^+\to\pi^0e^+\nu_e$ decays together with the lattice constraints, resulting in a  $|V_{\mathrm{ub}}|$ precision of approximately 11$\%$. The post-fit distributions of the partial branching fractions are illustrated in Figures \ref{fig:vubB0e}, \ref{fig:vubBpe} and \ref{fig:vubcombined}, for the individual fits to $B^0\to\pi^-e^+\nu_e$ and $B^+\to\pi^0e^+\nu_e$ and the combined fit, respectively.  The fitted values of $|V_{\mathrm{ub}}|$ are listed in Table \ref{table:vubresults}, together with the $\chi^2$ over the two degrees of freedom returned from each fit.

\begin{table} [h!]
\caption{Measured partial branching fractions of $B^0 \to \pi^- e^+ \nu_e$ decays in three bins of $q^2$, using 189.3\invfb of data, compared with the current world averages. The signal efficiencies derived from simulation are also listed.}\label{table:BFpiB0}
\def\arraystretch{1.2}%
\begin{tabular}{cccc}
\hline\hline
$q^2$ bin &  Signal efficiency & Unfolded signal yield & $\Delta \mathcal{B}$ \\\hline
\multicolumn{4}{c}{$B^0 \to \pi^- e^+ \nu_e$}\\\hline
0 $ \leq q^2 < 8$ GeV$^2$  & (0.189 $\pm$ 0.002)$\%$ & 15.5 $\pm$ 4.6  & (0.61 $\pm$ 0.18(stat) $\pm$ 0.03(syst)) $\times 10^{-4}$\\
8 $ \leq q^2 < 16$ GeV$^2$  & (0.239 $\pm$ 0.003)$\%$ & 15.3 $\pm$ 4.8  & (0.48 $\pm$ 0.15(stat) $\pm$ 0.02(syst)) $\times 10^{-4}$ \\
16 $ \leq q^2 \leq 26.4$ GeV$^2$ & (0.229 $\pm$ 0.003)$\%$ & 10.3 $\pm$ 4.2  & (0.34 $\pm$ 0.14(stat) $\pm$ 0.02(syst)) $\times 10^{-4}$\\\hline
Sum & -- &  41.1 $\pm$ 7.8 & (1.43 $\pm$ 0.27(stat) $\pm$ 0.07(syst)) $\times 10^{-4}$\\\hline
Fit over full $q^2$ range & (0.217 $\pm$ 0.002)$\%$ & 42.0 $\pm$  7.9 & (1.45 $\pm$ 0.27(stat) $\pm$ 0.07(syst)) $\times 10^{-4}$\\\hline
World average \cite{Zyla:2020zbs} & -- & -- & (1.50 $\pm$ 0.06) $\times 10^{-4}$\\\hline\hline

\end{tabular}
\end{table}

\begin{table} [h!]
\caption{Measured partial branching fractions of $B^+ \to \pi^0 e^+ \nu_e$ decays in three bins of $q^2$, using 189.3\invfb of data, compared with the current world averages. The signal efficiencies derived from simulation are also listed.}\label{table:BFpiBp}
\def\arraystretch{1.2}%
\begin{tabular}{cccc}
\hline\hline
$q^2$ bin &  Signal efficiency & Unfolded signal yield & $\Delta \mathcal{B}$ \\\hline
\multicolumn{4}{c}{$B^+ \to \pi^0 e^+ \nu_e$}\\\hline
0 $ \leq q^2 < 8$ GeV$^2$  & (0.329 $\pm$ 0.004)$\%$ & 12.9 $\pm$ 4.7  & (2.90 $\pm$ 1.12(stat) $\pm$ 0.19(syst)) $\times 10^{-5}$\\
8 $ \leq q^2 < 16$ GeV$^2$ & (0.439 $\pm$ 0.005)$\%$ & 18.1 $\pm$ 5.1  & (3.05 $\pm$ 0.91(stat) $\pm$ 0.20(syst)) $\times 10^{-5}$ \\
16 $ \leq q^2 \leq 26.4$ GeV$^2$ & (0.451 $\pm$ 0.006)$\%$ & 14.5 $\pm$ 4.9  & (2.38 $\pm$ 0.85(stat) $\pm$ 0.16(syst)) $\times 10^{-5}$\\\hline
Sum & -- &  45.5 $\pm$ 8.5  & (8.33 $\pm$ 1.67(stat) $\pm$ 0.55(syst)) $\times 10^{-5}$\\\hline
Fit over full $q^2$ range & (0.402 $\pm$ 0.003)$\%$ & 43.9 $\pm$ 8.3  & (8.06 $\pm$ 1.62(stat) $\pm$ 0.53(syst)) $\times 10^{-5}$\\\hline
World average \cite{Zyla:2020zbs} & -- & -- & (7.80 $\pm$ 0.27) $\times 10^{-5}$\\\hline\hline

\end{tabular}
\end{table}

\begin{table} [h!]
\caption{Fitted $|V_{\mathrm{ub}}|$ values from $\chi^2$ fits to the partial branching fraction distributions of $B^0\to\pi^-e^+\nu_e$ and $B^+\to\pi^0e^+\nu_e$ decays. The $\chi^2$ returned from the fit divided by the number of degrees of freedom is also listed.}\label{table:vubresults}
\begin{tabular}{ccc}
\hline\hline
Decay mode &   Fitted $|V_{\mathrm{ub}}|$ & Fit $\chi^2$/DOF \\
\hline
$B^0\to\pi^- e^+\nu_e$ & (3.71 $\pm$ 0.55) $\times 10^{-3}$ & 0.16\\
$B^+\to\pi^0 e^+\nu_e$ & (4.21 $\pm$ 0.63) $\times 10^{-3}$ & 0.02 \\
Combined fit & (3.88 $\pm$ 0.45) $\times 10^{-3}$ & 0.32 \\
 \hline\hline
\end{tabular}
\end{table}

\section{Systematic Uncertainties}
\label{sec:systematics}

A number of sources of systematic uncertainty are identified for this analysis and evaluated for the branching fraction measurements. The relative uncertainties for each source, in percent, are summarised in Table \ref{table:syserrors}. 

The uncertainties on the FEI calibration factors are determined from fits to the electron momentum spectrum of $B\to X e \nu_e$ decays. Sources of uncertainty in this fit include uncertainties on both the branching fractions and form factors of the various semileptonic components of $B\to X e \nu_e$, the lepton ID efficiency and fake rate uncertainties, tracking uncertainties and statistical uncertainties in the MC template distribution. The relative uncertainty on the calibration factor forms the dominant source of systematic uncertainty for the $B^0 \to \pi^- e^+ \nu_e$ analysis. For $B^+ \to \pi^0 e^+ \nu_e$, the relative uncertainty on the scaling factor to correct the $\pi^0$ efficiency forms the dominant source of systematic uncertainty, and is derived via an independent study of $D^0\to K^-\pi^+\pi^0$ and $D^0\to K^-\pi^+$ decays. 

The uncertainty on the number of $B\bar{B}$ events in the present data set includes systematic effects due to uncertainties on the luminosity, beam energy spread and shift, tracking efficiency and the selection efficiency of $B\bar{B}$ events. We represent the uncertainty on the signal reconstruction efficiency with a binomial error dependent on the size of the MC samples used for the analysis. We also combine the uncertainties on the world averages for the branching fractions $\mathcal{B}(\Upsilon$(4S)$\to B^+B^-$) and $\mathcal{B}(\Upsilon$(4S) $\to B^0\bar{B}^0$) and calculate the relative uncertainty on the fraction $f_{\mathrm{+0}}$. With regards to particle tracking, we assign a constant systematic uncertainty of 0.3$\%$ for each charged particle. For decay modes with multiple tracks we assume the associated uncertainties to be completely correlated. 

In addition, the electron efficiencies and pion and kaon fake rates are evaluated in bins of the electron momentum and polar angle, each with statistical and systematic uncertainties. The effect of these uncertainties on the signal reconstruction efficiency is determined through generating 200 variations on the nominal correction weights via Gaussian smearing. The relative uncertainty is then taken from the spread on the values of the reconstruction efficiency over all variations. The charged pion efficiencies and kaon fake rates are similarly evaluated in bins of the pion momentum $p$ and the polar angle $\theta$, as is done for the electron identification corrections. The relative systematic uncertainties are likewise determined via evaluating the effect of Gaussian smearing on the signal reconstruction efficiency, using 200 variations on the nominal correction weights.

\begin{table}
\centering
\caption{Sources of systematic uncertainty quoted as a percentage of the measured branching fractions. The statistical uncertainties are also listed.}\label{table:syserrors}
\begin{tabular}{cp{4em}p{4em}p{4em}p{4em}p{4em}p{4em}}
\hline\hline
Source                                & \multicolumn{3}{c}{$\%$  of}                     & \multicolumn{3}{c}{$\%$  of}       \\
~                                     & \multicolumn{3}{c}{$\mathcal{B}$($B^0 \to \pi^- e^+ \nu_e$)}                   & \multicolumn{3}{c}{$\mathcal{B}$($B^+ \to \pi^0 e^+ \nu_e$)}       \\
\hline
$q^2$ bin index & 1 & 2 & 3 & 1 & 2 & 3 \\\hline
 $N_{B\bar{B}}$                       & \multicolumn{6}{c}{2.9}\\  
 $f_{\mathrm{+0}}$                    & \multicolumn{6}{c}{1.2}\\  
 FEI calibration                      & \multicolumn{3}{c}{3.2}                               & \multicolumn{3}{c}{3.1}     \\                   
Tracking                             & \multicolumn{3}{c}{0.6}                      & \multicolumn{3}{c}{0.3} \\    
$\pi^0$ efficiency                   & \multicolumn{3}{c}{--}           & \multicolumn{3}{c}{4.8}  \\ 
Signal efficiency $\epsilon$ & 1.3  & 1.2 &  1.4                             & 1.3    & 1.2 &  1.3      \\ 
 Electron ID                            & 1.0  & 0.4 & 0.4                              & 1.0   & 0.5 & 0.5        \\       
 Pion ID                              & 0.4 & 0.4 & 0.4                                    & \multicolumn{3}{c}{--} \\                    
 \hline
 Total                                & 4.8  & 4.7 & 4.8                          & 6.7  & 6.7 & 6.7        \\\hline
 Stat. uncertainty & 29.5 & 31.3 & 41.2 & 38.6 & 29.8 & 35.7 \\
 \hline\hline
\end{tabular}
\end{table}

For $B \to \pi \ell^+ \nu_\ell$ decays, the systematic uncertainties from the modeling of  $B \to X_u \ell^+ \nu_\ell$ are expected to be small compared to other systematic uncertainties. Additional systematic uncertainties on the efficiencies of various selection criteria are not included, as these are expected to be negligibly small in comparison to other systematic effects.

\section{Summary}
\label{sec:conclusions}
We present an analysis of the semileptonic decays $B^0 \to \pi^- e^+ \nu_e$ and $B^+ \to \pi^0 e^+ \nu_e$ via hadronic tagging in a Belle II data sample corresponding to 189.3\invfb. We quote total branching fractions of $\mathcal{B}(B^0 \to \pi^- e^+ \nu_e)$ = (1.43 $\pm$ 0.27(stat) $\pm$ 0.07(syst)) $\times 10^{-4}$ and $\mathcal{B}(B^+ \to \pi^0 e^+ \nu_e)$ = (8.33 $\pm$ 1.67(stat) $\pm$ 0.55(syst)) $\times 10^{-5}$, based on the sum of the partial branching fractions in three bins of the momentum transfer to the leptonic system, $q^2$. These results are in agreement with the current world averages \cite{Zyla:2020zbs}. We extract a first Belle II determination of the CKM-matrix element $|V_{\mathrm{ub}}|$ via the hadronically tagged approach, with $|V_{\mathrm{ub}}|$ = (3.88 $\pm$ 0.45) $\times 10^{-3}$.

\clearpage
\section*{Acknowledgements}
\label{sec:acknowledgements}
We thank the SuperKEKB group for the excellent operation of the
accelerator; the KEK cryogenics group for the efficient
operation of the solenoid; and the KEK computer group for
on-site computing support.

\bibliography{belle2}

\providecommand{\href}[2]{#2}\begingroup\raggedright\begin{thebibliography}{10}

\bibitem{Keck:2018lcd}
T.~Keck et al., {\em {The Full Event Interpretation -- An exclusive tagging
  algorithm for the Belle II experiment}\/},
\href{http://dx.doi.org/10.1007/s41781-019-0021-8}{Comput. Softw. Big Sci. {\bf
  3} (2019)  6}.

\bibitem{Zyla:2020zbs}
P.~Zyla et al., {Particle Data Group}, {\em {Review of Particle Physics}\/},
  \href{http://dx.doi.org/10.1093/ptep/ptaa104}{PTEP {\bf 2020} (2020) no.~8,
  083C01}.

\bibitem{old:pilnu}
F.~Abudinen et al., {The Belle II Collaboration}, {\em {Exclusive $B\to
  X_u\ell\nu_\ell$ Decays with Hadronic Full-event-interpretation Tagging in
  62.8 fb$^{-1}$ of Belle II Data}\/},  2021.
\newblock \texttt{arXiv:2111.00710[hep-ex]}.

\bibitem{Abe:2010sj}
T.~Abe, {Belle II Collaboration}, {\em {Belle II Technical Design Report}\/},
\href{http://arxiv.org/abs/1011.0352}{{\tt arXiv:1011.0352 [physics.ins-det]}}.

\bibitem{Ramirez:1990db}
C.~Ramirez, J.~Donoghue, and G.~Burdman, {\em {Semileptonic $b \to u$
  Decay}\/},
\href{http://dx.doi.org/10.1103/PhysRevD.41.1496}{Phys. Rev. {\bf D41} (1990)
  1496--1503}.

\bibitem{Lange:2005ll}
B.~O. Lange, M.~Neubert, and G.~Paz, {\em {Theory of Charmless Inclusive B
  Decays and the Extraction of $V_{\mathrm{ub}}$}\/},
\href{http://dx.doi.org/10.1103/PhysRevD.72.073006}{Phys. Rev. {\bf D72} (2005)
   073006}.

\bibitem{markus_prim_2020_3965699}
M.~Prim, {\em b2-hive/eFFORT v0.1.0\/},
  {https://doi.org/10.5281/zenodo.3965699}, July, 2020.

\bibitem{Keck:2016tk}
T.~Keck, {\em {FastBDT: A Speed-Optimized Multivariate Classification Algorithm
  for the Belle II Experiment}\/},
\href{http://dx.doi.org/10.1007/s41781-017-0002-8}{Comput. Softw. Big Sci. {\bf
  1} (2017)  2}.

\bibitem{belle2tracking:2021}
{Belle II Tracking Group}, {\em Track finding at Belle II\/},
  \href{http://dx.doi.org/https://doi.org/10.1016/j.cpc.2020.107610}{Comp.
  Phys. Comm. {\bf 259} (2021)  107610}.

\bibitem{Sibidanov:2013sb}
A.~Sibidanov et al., {Belle Collaboration}, {\em {Study of Exclusive $B \to X_u
  \ell \nu$ Decays and Extraction of $|V_{\mathrm{ub}}|$ using Full
  Reconstruction Tagging at the Belle Experiment}\/},
Phys. Rev. {\bf D88} (2013)  032005.

\bibitem{Wolfram:1978fw}
G.~C. Fox and S.~Wolfram, {\em {Observables for the Analysis of Event Shapes in
  $e^+e^-$ Annihilation and Other Processes}\/},
\href{http://dx.doi.org/https://doi.org/10.1103/PhysRevLett.41.1581}{Phys. Rev.
  Lett. {\bf 41} (1978)  1581}.

\bibitem{Kuhr:B2}
T.~Kuhr et al., {\em {The Belle II Core Software}\/},
\href{http://dx.doi.org/10.1007/s41781-018-0017-9}{Comput. Softw. Big Sci. {\bf
  3} (2019)  1}.

\bibitem{LeptonID:2318}
{LeptonID group and Belle II Collaboration}, {\em Muon and electron
  identification efficiencies and hadron-lepton mis-identification rates at
  Belle II for Moriond 2021\/},
\newblock Mar, 2021.
\newblock BELLE2-CONF-PH-2021-002.

\bibitem{cern:roounfold}
T.~Adye, {\em Unfolding Algorithms and Tests using RooUnfold\/},  2011.
\newblock \texttt{arXiv:1105.1160 [physics.data-an]}.

\bibitem{Bailey:2015bl}
J.~A. Bailey et al., {\em {$|V_{\mathrm{ub}}|$ from $B\to\pi\ell\nu$ decays and
  $2 + 1$-flavor lattice QCD}\/},
  \href{http://dx.doi.org/10.1103/PhysRevD.92.014024}{Phys. Rev. {\bf D92}
  (2015)  014024}.

\bibitem{BCL}
C.~Bourrely, L.~Lellouch, and I.~Caprini, {\em Model-independent description of
  $B\ensuremath{\rightarrow}\ensuremath{\pi}l\ensuremath{\nu}$ decays and a
  determination of $|{V}_{ub}|$\/},
  \href{http://dx.doi.org/10.1103/PhysRevD.79.013008}{Phys. Rev. {\bf D79}
  (2009)  013008}.

\end{thebibliography}\endgroup
\bibliographystyle{belle2-note}
\clearpage
\appendix

\section{Partial branching fraction covariance matrices}\label{sec:PBFcovmatrices}
The total covariance matrices for each set of partial branching fractions are built from the corresponding statistical and systematic uncertainties. We assume each systematic effect is completely independent from all others and from the statistical uncertainties, and thus we obtain the total covariance matrices by adding the individual matrices for each given uncertainty. Statistical uncertainties between $q^2$ bins are taken to be completely uncorrelated and thus their covariance matrices are diagonal. We conservatively assume each systematic uncertainty to be completely correlated between $q^2$ bins.\\

\begin{table} [h!]
\caption{The covariance matrices for each set of partial branching fractions derived from $B^0\to\pi^-e^+\nu_e$ and $B^+\to\pi^0 e^+\nu_e$ decays.}\label{table:PBF_covmatrices}
\begin{tabular}{c|ccc}\hline\hline
\multicolumn{3}{c}{$B^0\to\pi^-e^+\nu_e$} & ($\times 10^{-8}$)\\\hline

$q^2$ bin index & 1 & 2 & 3\\\hline
1 & 0.033278 &  &  \\
2 & 0.000668 & 0.022883 &  \\
3 & 0.000478 & 0.000368 & 0.019015 \\\hline
\end{tabular}

\begin{tabular}{c|ccc}
\multicolumn{3}{c}{$B^+\to\pi^0 e^+\nu_e$} & ($\times 10^{-10}$)\\\hline

$q^2$ bin index & 1 & 2 & 3\\\hline
1 & 1.265127 &  &  \\
2 & 0.038937 & 0.862037 &  \\
3 & 0.030478 & 0.031726 & 0.749881 \\\hline\hline
\end{tabular}

\end{table}
\clearpage 

\section{$\chi^2$ fits, further details}\label{sec:vubextra}

A number of extra details related to the $\chi^2$ fits to the distributions of partial branching fractions are provided.

\begin{table} [h!]
\caption{The pre- and post-fit BCL parameters from the $\chi^2$ fits to the partial branching fractions derived from $B^0\to\pi^-e^+\nu_e$ and $B^+\to\pi^0 e^+\nu_e$ decays.}\label{table:vubfitparams}
\begin{tabular}{c|ccccccccc}\hline\hline
\multicolumn{9}{c}{Input parameters}\\\hline
& b0p & b1p & b2p & b3p & b00 & b10 & b20 & b30  \\\hline
 & 0.407 & $-$0.65 & $-$0.46 & 0.4 & 0.507 & $-$1.77 & 1.27 & 4.2   \\\hline
\multicolumn{9}{c}{Input correlation matrix}\\\hline
& b0p & b1p & b2p & b3p & b00 & b10 & b20 & b30\\\hline
b0p & 1 &  &  &  &  &  &  &  \\
b1p & 0.451 & 1 &  &  &  &  &  &  \\
b2p & 0.161 & 0.757 & 1 &  &  &  &  &  \\
b3p & 0.102 & 0.665 & 0.988 & 1 &  &  &  &  \\
b00 & 0.331 & 0.430 & 0.482 & 0.484 & 1 &  &  &  \\
b10 & 0.346 & 0.817 & 0.847 & 0.833 & 0.447 & 1 &  &  \\
b20 & 0.292 & 0.854 & 0.951 & 0.913 & 0.359 & 0.827 & 1 &  \\
b30 & 0.216 & 0.699 & 0.795 & 0.714 & 0.189 & 0.500 & 0.838 & 1 \\
\end{tabular}
\footnotesize
\begin{tabular}{p{7em}|p{6em}p{6em}p{6em}p{6em}p{6em}p{6em}p{1em}p{1em}}\hline\hline
\multicolumn{9}{c}{\normalsize Fit parameters}\\\hline
& b0p & b1p & b2p & b3p & b00 & b10 &  &  \\\hline
$B^0\to\pi^- e^+\nu_e$ & 0.41 $\pm$ 0.01 & $-$0.58 $\pm$ 0.10 & $-$0.05 $\pm$ 0.50 & 0.98 $\pm$ 0.80 & 0.51 $\pm$ 0.02 & $-$1.69 $\pm$ 0.12 & & \\
$B^+\to\pi^0 e^+\nu_e$ & 0.41 $\pm$ 0.01 & $-$0.63 $\pm$ 0.10 & $-$0.35 $\pm$ 0.47 & 0.56 $\pm$ 0.77 & 0.51 $\pm$ 0.02 & $-$1.75 $\pm$ 0.11 & & \\
Combined fit & 0.41 $\pm$ 0.01 & $-$0.60 $\pm$ 0.09 & $-$0.15 $\pm$ 0.40 & 0.84 $\pm$ 0.68 & 0.51 $\pm$ 0.02 & $-$1.71 $\pm$ 0.10 & & \\\hline\hline
\end{tabular}
\footnotesize
\begin{tabular}{p{7em}|p{6em}p{6em}p{10em}p{6em}p{6em}p{1em}p{1em}p{2em}}
& b20 & b30 & &  & & & &  \\\hline
$B^0\to\pi^- e^+\nu_e$ &  1.65 $\pm$ 0.46  & 4.74 $\pm$ 1.03 & & & & & &   \\
$B^+\to\pi^0 e^+\nu_e$ & 1.38 $\pm$ 0.43 & 4.36 $\pm$ 1.01 & & & &  & &  \\
Combined fit & 1.56 $\pm$ 0.37 & 4.61 $\pm$ 0.96 & & & &  & &  \\\hline\hline

\end{tabular}

\end{table}

\begin{table} [h!]
\caption{The post-fit covariance matrices from the $\chi^2$ fits to the partial branching fractions derived from $B^0\to\pi^-e^+\nu_e$ and $B^+\to\pi^0 e^+\nu_e$ decays.}\label{table:vubfitcovmatrices}
\begin{tabular}{c|ccccccccc}\hline\hline
\multicolumn{9}{c}{$B^0\to\pi^- e^+\nu_e$ } & ($\times 10^{-4}$)\\\hline
& $|V_{\mathrm{ub}}|$ & b0p & b1p & b2p & b3p & b00 & b10 & b20 & b30 \\\hline
 $|V_{\mathrm{ub}}|$ &  0.0031 &  &  & 
 &  &  &  & 
 & \\
b0p & $-$0.0296 & 2.0952 &  & 
 &  &  &  & 
 &  \\
b1p & $-$0.3461 & 5.9133 & 109.4110 & 
 &   &  &  & 
 &  \\
b2p & $-$1.5366 & $-$7.7452 & 182.5664 & 2470.5861
 &   &  &  & 
 &  \\
b3p & $-$2.0388 & $-$21.0003 & 134.9172 & 3786.0401
 &  6345.9003 &  &  & 
 &  \\
b00 & $-$0.0278 & 0.7135 & 3.6421 & 24.3506
 &  40.8809 & 3.9682 &  & 
 & \\
b10 & $-$0.3473 & 3.8461 & 68.6689 & 342.3118
 &  535.8826 & 4.7152 & 137.4099 & 
 & \\
b20 & $-$1.5971 & 8.6476 & 292.3029 & 1890.2535
 &  2699.2709 & 0.4659 & 285.3763 & 2092.5181
 &  \\
b30 & $-$2.2303 & 7.3812 & 412.8882 & 2875.0134
 &  3207.5347 & $-$31.6807 & $-$42.7719 & 3131.5516
 &  10679.612\\\hline
\multicolumn{9}{c}{$B^+\to\pi^0 e^+\nu_e$} & ($\times 10^{-4})$\\\hline
& $|V_{\mathrm{ub}}|$ & b0p & b1p & b2p & b3p & b00 & b10 & b20 & b30 \\\hline
$|V_{\mathrm{ub}}|$  & 0.0039 &  &  & 
&  &  &  & 
 & \\
b0p &  $-$0.0329 & 2.0766 &  & 
 &  &  &  & 
 &  \\
b1p &  $-$0.3523 & 5.5307 & 101.7664 & 
 &   &  &  & 
 &  \\
b2p & $-$1.4964 & $-$9.7706 & 141.3677 & 2243.5309
 &   &  &  & 
 &  \\
b3p &  $-$1.9627 & $-$23.7964 & 77.8032 & 3469.7625
 &  5904.8975 &  &  & 
 &  \\
b00 &  $-$0.0285 & 0.6846 & 3.0501 & 21.1118
 &  36.3769 & 3.9220 &  & 
 & \\
b10 &  $-$0.3486 & 3.4395 & 60.4387 & 297.3735
 &  473.4102 & 4.0727 & 128.4830 & 
 & \\
b20 &  $-$1.5873 & 6.6995 & 252.8578 & 1674.3876
 &  2399.0297 & $-$2.6195 & 242.5307 & 1886.8432
 &  \\
b30 &  $-$2.2114 & 4.6325 & 357.3143 & 2571.0205
 &  2784.7566 & $-$36.0278 & $-$103.1253 & 2841.8612
 &  10271.6000\\\hline
 \multicolumn{9}{c}{Combined fit} & ($\times 10^{-4}$)\\\hline
 & $|V_{\mathrm{ub}}|$ & b0p & b1p & b2p & b3p & b00 & b10 & b20 & b30 \\\hline
$|V_{\mathrm{ub}}|$ & 0.0020 &  &  & 
 &  &  &  & 
 & \\
b0p & $-$0.0271 & 2.0618 &  & 
 &  &  &  & 
 &  \\
b1p & $-$0.2356 & 5.0730 & 85.0987 & 
 &  &  &  & 
 &  \\
b2p & $-$0.8452 & $-$12.2496 & 40.3805 & 1599.9840
 &   &  &  & 
 &  \\
b3p & $-$1.0540 & $-$27.2303 & $-$65.2311 & 2549.9842
 &  4588.2605 &  &  & 
 &  \\
b00 & $-$0.0193 & 0.6524 & 1.7764 & 13.1221
 &  24.9894 & 3.8224 &  & 
 & \\
b10 & $-$0.2204 & 2.9559 & 41.6042 & 179.7174
 &  305.8415 & 2.6032 & 106.8052 & 
 & \\
b20 & $-$0.9705 & 4.3511 & 160.3458 & 1093.3470
 &  1570.7014 & $-$9.8650 & 135.7003 & 1360.0744
 &  \\
b30 & $-$1.3414 & 1.2991 & 226.1362 & 1747.3173
 &  1610.5369 & $-$46.3005 & $-$254.5869 & 2095.0420
 &  9212.8063 \\
 \hline\hline
\end{tabular}
\end{table}

\end{document}